\PassOptionsToPackage{unicode}{hyperref}
\PassOptionsToPackage{hyphens}{url}
\PassOptionsToPackage{dvipsnames,svgnames,x11names}{xcolor}
\documentclass[
]{article}
\usepackage{xcolor}
\usepackage{amsmath,amssymb}
\usepackage{iftex}
\ifPDFTeX
  \usepackage[T1]{fontenc}
  \usepackage[utf8]{inputenc}
  \usepackage{textcomp} 
\else 
  \usepackage{unicode-math} 
  \defaultfontfeatures{Scale=MatchLowercase}
  \defaultfontfeatures[\rmfamily]{Ligatures=TeX,Scale=1}
\fi
\usepackage{lmodern}
\ifPDFTeX\else
\fi
\IfFileExists{upquote.sty}{\usepackage{upquote}}{}
\IfFileExists{microtype.sty}{
  \usepackage[]{microtype}
  \UseMicrotypeSet[protrusion]{basicmath} 
}{}
\makeatletter
\@ifundefined{KOMAClassName}{
  \IfFileExists{parskip.sty}{%
    \usepackage{parskip}
  }{
    \setlength{\parindent}{0pt}
    \setlength{\parskip}{6pt plus 2pt minus 1pt}}
}{
  \KOMAoptions{parskip=half}}
\makeatother
\makeatletter
\ifx\paragraph\undefined\else
  \let\oldparagraph\paragraph
  \renewcommand{\paragraph}{
    \@ifstar
      \xxxParagraphStar
      \xxxParagraphNoStar
  }
  \newcommand{\xxxParagraphStar}[1]{\oldparagraph*{#1}\mbox{}}
  \newcommand{\xxxParagraphNoStar}[1]{\oldparagraph{#1}\mbox{}}
\fi
\ifx\subparagraph\undefined\else
  \let\oldsubparagraph\subparagraph
  \renewcommand{\subparagraph}{
    \@ifstar
      \xxxSubParagraphStar
      \xxxSubParagraphNoStar
  }
  \newcommand{\xxxSubParagraphStar}[1]{\oldsubparagraph*{#1}\mbox{}}
  \newcommand{\xxxSubParagraphNoStar}[1]{\oldsubparagraph{#1}\mbox{}}
\fi
\makeatother

\usepackage{color}
\usepackage{fancyvrb}

\DefineVerbatimEnvironment{Highlighting}{Verbatim}{commandchars=\\\{\}}
\usepackage{framed}
\definecolor{shadecolor}{RGB}{241,243,245}
\newenvironment{Shaded}{\begin{snugshade}}{\end{snugshade}}

\newcommand{\AttributeTok}[1]{\textcolor[rgb]{0.40,0.45,0.13}{#1}}

\newcommand{\CommentTok}[1]{\textcolor[rgb]{0.37,0.37,0.37}{#1}}

\newcommand{\ConstantTok}[1]{\textcolor[rgb]{0.56,0.35,0.01}{#1}}
\newcommand{\ControlFlowTok}[1]{\textcolor[rgb]{0.00,0.23,0.31}{\textbf{#1}}}

\newcommand{\DecValTok}[1]{\textcolor[rgb]{0.68,0.00,0.00}{#1}}

\newcommand{\FloatTok}[1]{\textcolor[rgb]{0.68,0.00,0.00}{#1}}
\newcommand{\FunctionTok}[1]{\textcolor[rgb]{0.28,0.35,0.67}{#1}}

\newcommand{\NormalTok}[1]{\textcolor[rgb]{0.00,0.23,0.31}{#1}}

\newcommand{\OtherTok}[1]{\textcolor[rgb]{0.00,0.23,0.31}{#1}}

\newcommand{\SpecialCharTok}[1]{\textcolor[rgb]{0.37,0.37,0.37}{#1}}

\newcommand{\StringTok}[1]{\textcolor[rgb]{0.13,0.47,0.30}{#1}}

\usepackage{longtable,booktabs,array}
\usepackage{calc} 
\usepackage{etoolbox}
\makeatletter
\patchcmd\longtable{\par}{\if@noskipsec\mbox{}\fi\par}{}{}
\makeatother
\IfFileExists{footnotehyper.sty}{\usepackage{footnotehyper}}{\usepackage{footnote}}
\makesavenoteenv{longtable}
\usepackage{graphicx}
\makeatletter
\newsavebox\pandoc@box
\newcommand*\pandocbounded[1]{
  \sbox\pandoc@box{#1}%
  \Gscale@div\@tempa{\textheight}{\dimexpr\ht\pandoc@box+\dp\pandoc@box\relax}%
  \Gscale@div\@tempb{\linewidth}{\wd\pandoc@box}%
  \ifdim\@tempb\p@<\@tempa\p@\let\@tempa\@tempb\fi
  \ifdim\@tempa\p@<\p@\scalebox{\@tempa}{\usebox\pandoc@box}%
  \else\usebox{\pandoc@box}%
  \fi%
}
\def\fps@figure{htbp}
\makeatother

\NewDocumentCommand\citeproctext{}{}

\makeatletter
 \let\@cite@ofmt\@firstofone
 \def\@biblabel#1{}
 \def\@cite#1#2{{#1\if@tempswa , #2\fi}}
\makeatother
\newlength{\cslhangindent}
\newlength{\csllabelwidth}
\newenvironment{CSLReferences}[2] 
 {\begin{list}{}{%
  \setlength{\itemindent}{0pt}
  \setlength{\leftmargin}{0pt}
  \setlength{\parsep}{0pt}
  \ifodd #1
   \setlength{\leftmargin}{\cslhangindent}
   \setlength{\itemindent}{-1\cslhangindent}
  \fi
  \setlength{\itemsep}{#2\baselineskip}}}
 {\end{list}}
\usepackage{calc}

\usepackage{arxiv}
\usepackage{orcidlink}
\usepackage{amsmath}
\makeatletter
\@ifpackageloaded{caption}{}{\usepackage{caption}}
\AtBeginDocument{%
\ifdefined\contentsname
  \renewcommand*\contentsname{Table of contents}
\else
  \newcommand\contentsname{Table of contents}
\fi
\ifdefined\listfigurename
  \renewcommand*\listfigurename{List of Figures}
\else
  \newcommand\listfigurename{List of Figures}
\fi
\ifdefined\listtablename
  \renewcommand*\listtablename{List of Tables}
\else
  \newcommand\listtablename{List of Tables}
\fi
\ifdefined\figurename
  \renewcommand*\figurename{Figure}
\else
  \newcommand\figurename{Figure}
\fi
\ifdefined\tablename
  \renewcommand*\tablename{Table}
\else
  \newcommand\tablename{Table}
\fi
}
\@ifpackageloaded{float}{}{\usepackage{float}}
\floatstyle{ruled}
\@ifundefined{c@chapter}{\newfloat{codelisting}{h}{lop}}{\newfloat{codelisting}{h}{lop}[chapter]}
\floatname{codelisting}{Listing}

\makeatother
\makeatletter
\@ifpackageloaded{caption}{}{\usepackage{caption}}
\@ifpackageloaded{subcaption}{}{\usepackage{subcaption}}
\makeatother
\usepackage{bookmark}
\IfFileExists{xurl.sty}{\usepackage{xurl}}{} 
\hypersetup{
  pdftitle={The Role of Congeniality in Multiple Imputation for Doubly Robust Causal Estimation},
  pdfauthor={Lucy D'Agostino McGowan},
  pdfkeywords={multiple imputation, inverse probability
weighting, doubly robust, causal inference, missing data},
  colorlinks=true,
  linkcolor={blue},
  filecolor={Maroon},
  citecolor={Blue},
  urlcolor={Blue},
  pdfcreator={LaTeX via pandoc}}

\newcommand{\runninghead}{A Preprint }
\title{The Role of Congeniality in Multiple Imputation for Doubly Robust
Causal Estimation}
\author{\textbf{Lucy D'Agostino McGowan}\\Department of Statistical
Sciences\\Wake Forest University, Winston-Salem, NC
27104\\\\\href{mailto:mcgowald@wfu.edu}{mcgowald@wfu.edu}}
\date{}
\begin{document}
\maketitle
\begin{abstract}
This paper provides clear and practical guidance on the specification of
imputation models when multiple imputation is used in conjunction with
doubly robust estimation methods for causal inference. Through
theoretical arguments and targeted simulations, we demonstrate that if a
confounder has missing data, the corresponding imputation model must
include all variables appearing in either the propensity score model or
the outcome model, in addition to both the exposure and the outcome, and
that these variables must enter the imputation model in the same
functional form as in the final analysis. Violating these conditions can
lead to biased treatment effect estimates, even when both components of
the doubly robust estimator are correctly specified. We present a
mathematical framework for doubly robust estimation combined with
multiple imputation, establish the theoretical requirements for proper
imputation in this setting, and demonstrate the consequences of
misspecification through simulation. Based on these findings, we offer
concrete recommendations to ensure valid inference when using multiple
imputation with doubly robust methods in applied causal analyses.
\end{abstract}
{\bfseries \emph Keywords}
\def\sep{\textbullet\ }
multiple imputation \sep inverse probability weighting \sep doubly
robust \sep causal inference \sep 
missing data

In observational studies attempting to estimate causal effects
researchers increasingly rely on methods that combine multiple analysis
models to estimate treatment effects. One common approach uses
propensity score models, which estimate the probability of treatment
assignment given observed covariates to help balance treatment and
control groups. Alternatively, researchers can fit outcome models that
directly relate observed covariates to the outcome distribution, relying
on the model specification to control for confounding. Doubly robust
estimators, which combine both approaches, have become increasingly
popular as they offer protection against model misspecification by
combining propensity score models with outcome regression models
(Scharfstein, Rotnitzky, and Robins 1999; Robins 2000; Lunceford and
Davidian 2004; Bang and Robins 2005; Kang and Schafer 2007). Targeted
maximum likelihood estimation (TMLE) has emerged as another doubly
robust approach that combines propensity score and outcome modeling
while offering additional efficiency properties (Van der Laan, Rose, et
al. 2011). Applied researchers frequently encounter missing data in
covariates or outcomes. While multiple imputation (MI) has emerged as a
common solution for handling such missingness under the missing at
random (MAR) assumption, there remains limited guidance on how to
properly specify imputation models when working with these combined
analysis approaches.

For handling propensity score methods with multiply imputed data, the
preferred strategy involves generating \(m\) imputed datasets,
estimating the propensity score and treatment effect in each one, and
then combining the \(m\) resulting average treatment effect estimates
using Rubin's rules, an approach sometimes termed the ``within''
approach in the literature. This approach has been shown to outperform
alternatives that, for example, pool propensity scores across
imputations before estimating treatment effects (Granger, Sergeant, and
Lunt 2019; Leyrat et al. 2019). However, while the general structure of
this approach is now well accepted, the accompanying studies offer
limited and inconsistent guidance on the specification of the imputation
model itself, particularly when multiple analysis models are involved as
is the case with doubly robust estimators.

The imputation model's specification becomes especially critical when
working with doubly robust estimators. These estimators incorporate both
a propensity score model and outcome model, and are consistent if either
model is correctly specified (Robins 2000; Lunceford and Davidian 2004).
This dual-model structure creates unique challenges for imputation model
specification. Existing methodological literature provides fragmented
guidance across different statistical frameworks, making systematic
implementation difficult for practitioners. In theory, the imputation
model must preserve the joint distribution of all variables that will be
used in any subsequent analysis models to ensure congeniality (Daniels,
Wang, and Marcus 2014; Bartlett and Hughes 2020). Congeniality refers to
the compatibility between the imputation model and the subsequent
analysis models, specifically, whether the imputation model adequately
captures the relationships among variables as they are included in the
analysis models. Failure to maintain this congeniality can introduce
bias even when the analysis models themselves are correctly specified.

A common oversight in applied research is the exclusion of important
variables or their functional forms from imputation models. For example,
the simulation study by Mitra and Reiter (Mitra and Reiter 2016), which
aimed to compare MI--propensity score integration strategies, excluded
the outcome variable from the imputation model entirely. As noted by
Penning de Vries and Groenwold Penning de Vries and Groenwold (2016),
this decision invalidates the results and conclusions of that
comparison, because excluding the outcome from the imputation model
leads to biased estimates (D'Agostino McGowan, Lotspeich, and Hepler
2024). Similarly, Williamson et al. (Williamson, Forbes, and Wolfe 2012)
examine multiple imputation under several modeling strategies but
incorrectly excluded the exposure variable from their imputation model
(which included only confounders and the outcome). The authors
attributed their suboptimal MI results to poor model fit, while the
likely underlying issue stemmed from lack of congeniality between their
imputation and analysis models. This pattern of observing bias when
comparing results to MI methods without explicitly connecting it to
congeniality principles appears repeatedly in the literature. In another
recent example, Dashti et al. Dashti et al. (2024) compared various MI
scenarios paired with TMLE and identified several scenarios where
default MI methods yielded biased results while methods that matched the
complexity of the analysis models performed better. While their
observations were likely fundamentally driven by compatibility between
imputation and analysis models, the connection to the broader principle
of congeniality and its systematic application across different doubly
robust frameworks was not made explicit. Collectively, these examples
illustrate the need for systematic guidance that makes the connection
between theoretical congeniality principles and practical implementation
explicit. While we highlight these specific studies, it is important to
note that they are commendable for providing sufficient methodological
detail to allow readers to evaluate their imputation models, a level of
transparency that is unfortunately rare in clinical research, suggesting
that such misspecification issues are likely far more widespread than
the published literature would indicate. When estimating causal effects,
intuition might suggest that certain variables could be safely omitted
from imputation models, such as precision variables that affect only the
outcome. However, when a confounder has missing values these variables
must be included in the corresponding imputation models when they appear
in subsequent analyses, and their inclusion must match the functional
form in the subsequent models.

This paper provides theoretical and empirical guidance on what must be
included in the imputation model when MI is combined with doubly robust
estimation methods, focusing on settings with binary treatment and
continuous pre-exposure covariates and outcome. We show through
theoretical argument and simulations that when using MI in conjunction
with propensity score estimation and outcome modeling, the imputation
model must include all variables used in either the propensity score
model or the outcome model(s), and these variables must appear in the
same functional form as they do in the subsequent analyses (Von Hippel
2009; White, Royston, and Wood 2011; Hippel 2013). We demonstrate that
violating these conditions leads to biased estimates of treatment
effects, even when both components of the doubly robust estimator are
correctly specified. Our contribution consolidates theoretical
requirements into practical implementation guidelines while quantifying
the empirical consequences of specification violations.

We organize the remainder of this paper as follows. First, we present
the mathematical framework for doubly robust estimation and multiple
imputation. Next, we develop the theoretical requirements for proper
imputation model specification in this context, with particular
attention to the challenges posed by combining multiple analysis models.
We then use targeted simulations to illustrate how violations of these
principles affect bias and efficiency. Finally, we offer practical
recommendations for analysts using MI alongside doubly robust estimators
in applied causal inference settings.

\section{Mathematical Framework}\label{mathematical-framework}

We consider a standard potential outcomes framework for causal
inference. Let \(X \in \{0,1\}\) denote a binary exposure,
\(Y \in \mathbb{R}\) an outcome, and \(\mathbf{Z} \in \mathbb{R}^p\) a
vector of pre-exposure covariates. For each individual, let \(Y(1)\) and
\(Y(0)\) denote the potential outcomes under treatment and control,
respectively. Our estimand is the average treatment effect (ATE),

\[
\Delta = \mathbb{E}[Y(1) - Y(0)].
\]

\begin{figure}[H]

\centering{

\pandocbounded{\includegraphics[keepaspectratio]{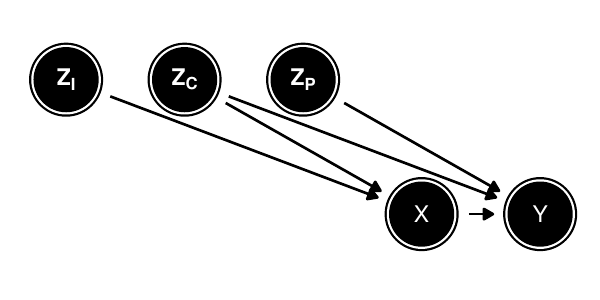}}

}

\caption{\label{fig-dag}Directed acyclic graph describing the
relationship between \(X\), \(Y\), and \(\mathbf{Z}\) where
\(\mathbf{Z}_I\) represent instruments, \(\mathbf{Z}_C\) confounders,
and \(\mathbf{Z}_P\) precision variables.}

\end{figure}

To clarify the causal roles of covariates, we partition the covariate
vector \(\mathbf{Z}\) into three disjoint subsets: \[
\mathbf{Z} = (\mathbf{Z}_{I},\, \mathbf{Z}_{P},\, \mathbf{Z}_{C}),
\]

where \(\mathbf{Z}_{I}\) denotes instruments (variables that causally
affect \(X\) but not \(Y\)), \(\mathbf{Z}_{P}\) denotes precision
variables (variables that causally affect \(Y\) but not \(X\)), and
\(\mathbf{Z}_{C}\) denotes confounders (variables that causally affect
both \(X\) and \(Y\). Figure~\ref{fig-dag} is a directed acyclic graph
(DAG) that displays these relationships.

\phantomsection\label{cell-fig-dag}

To identify \(\Delta\), we rely on the following standard assumptions:
(1) Consistency: \(Y = Y(X)\), i.e., the observed outcome corresponds to
the potential outcome under the observed exposure, (2) Exchangeability
(Unconfoundedness):
\(Y(1), Y(0) \perp\!\!\!\perp X \mid \mathbf{Z}_{C}\), and (3)
Positivity: \(0 < \mathbb{P}(X = 1 \mid \mathbf{Z}_{C}) < 1\).

Under these assumptions, the ATE can be identified using several
estimation strategies. One common approach is inverse probability
weighting (IPW), where we define the propensity score as \[
\pi(\mathbf{Z}_{C}) = \mathbb{P}(X = 1 \mid \mathbf{Z}_{C}),
\] and construct the IPW estimator as

\begin{equation}\phantomsection\label{eq-ipw}{
\hat{\Delta}_{\text{IPW}} = \frac{1}{n} \sum_{i=1}^n \left[ \frac{X_i Y_i}{{\pi}(\mathbf{Z}_{C,i})} - \frac{(1 - X_i) Y_i}{1 - {\pi}(\mathbf{Z}_{C,i})} \right].
}\end{equation}

We can optionally fit a doubly robust estimator, which combines the IPW
and outcome regression approaches. Often this involves fitting separate
outcome regression models for the exposed and unexposed groups, then
combining these results with inverse probability weights. While there
are several approaches to incorporating both propensity scores and
outcome regressions, in this paper we focus on augmented inverse
probability weighting Bang and Robins (2005). This doubly robust
estimator is defined as follows:

\begin{equation}\phantomsection\label{eq-dr}{
\hat{\Delta}_{\text{DR}} = \frac{1}{n} \sum_{i=1}^n \left[\mu_1(\mathbf{Z}_{C,i})-\mu_0(\mathbf{Z}_{C,i})\right] + \frac{1}{n} \sum_{i=1}^n\left[\frac{X_i\left\{Y_i-\mu_1(\mathbf{Z}_{C,i})\right\}}{\pi(\mathbf{Z}_{C,i})}-\frac{(1-X_i)\left\{Y_i-\mu_0(\mathbf{Z}_{C,i})\right\}}{1-\pi(\mathbf{Z}_{C,i})}\right],
}\end{equation}

where \(\mu_1(\mathbf{Z}_{C})\) and \(\mu_0(\mathbf{Z}_{C})\) are
predicted outcomes of a regression model fit among the exposed (\(X=1\))
and unexposed (\(X=0\)) respectively. Note if all the same variables are
included in both outcome models, fitting these models separately among
the exposed and unexposed is effectively including an interaction term
between all variables and the exposure.

An advantage of doubly robust estimators is that they remain consistent
if either the propensity score model (\(\pi\)) or the outcome models
(\(\mu\)) are correctly specified. Even when both models use the same
covariates \(\mathbf{Z}_{C}\), this property may be useful in that it
ensures consistency as long as at least one model correctly captures the
functional relationship between variables. It is also possible to
estimate \({\mu}\) using both \(\mathbf{Z}_{C}\) and \(\mathbf{Z}_{P}\),
as including variables that are predictive of the outcome (but not the
exposure) can improve the efficiency of the estimator without
introducing bias (Hahn 2004; Lunceford and Davidian 2004; Brookhart et
al. 2006; De Luna, Waernbaum, and Richardson 2011; Franklin et al. 2015;
Tang et al. 2023). In fact, Craycroft et al. Craycroft, Huang, and Kong
(2020) even recommend including both confounders (\(\mathbf{Z}_{C}\))
and precision variables (\(\mathbf{Z}_{P}\)) when estimating the
propensity score model \(\pi\). They also demonstrate that including
instruments (\(\mathbf{Z}_{I}\)) in the propensity score model is not
recommended.

\section{Missing Data and Multiple Imputation}\label{sec-mi}

Data may be missing in the outcome \(Y\) and/or covariates
\(\mathbf{Z}\). Let \(M^Y\) and \(M^{\mathbf{Z}}\) denote missingness
indicators, where \(M^V = 1\) if variable \(V \in \{Y, \mathbf{Z}\}\) is
missing. Let \(\mathbf{O} = (Y^*, \mathbf{Z}^*)\) represent the observed
data, where for any variable \(V\), \(V^* = V\) if \(M^V = 0\), and is
missing otherwise.

We assume that the data are MAR, meaning that the probability of
missingness depends only on the observed data. Specifically:

\[
\mathbb{P}(M^V = 1 \mid X, Y, \mathbf{Z}) = \mathbb{P}(M^V = 1 \mid \mathbf{O}).
\]

Under this MAR assumption, MI is a valid strategy for handling missing
data. MI proceeds by specifying a set of imputation models. These models
are used to generate \(m\) completed datasets by drawing from the
posterior predictive distribution of the missing values. For a variable
\(V\) with missingness, the imputation model aims to draw from its
posterior predictive distribution:

\[\mathbb{P}(V_{mis}|\mathbf{O}, \boldsymbol{\theta}_V)\] where
\(\boldsymbol\theta_V\) represents the parameters of the imputation
model for variable \(V\). For example, when imputing a continuous
confounder \(Z_C\) with missing values, a linear model might be
specified as:

\begin{equation}\phantomsection\label{eq-imp}{
Z_C = \beta_0+\beta_1X+\beta_2Z_I+\beta_3Z_p+\beta_4Y+\varepsilon,\text{      }\varepsilon\sim\mathcal{N}(0,1),
}\end{equation}

where draws from this model are used to generate stochastic imputations
in each multiply imputed dataset. Equation~\ref{eq-imp} assumes \(Z_I\)
and \(Z_P\) both appear in the subsequent analysis models (i.e.~the
propensity score and/or outcome models) in linear form. Of course, in
practice the imputation model can include nonlinear terms and/or
additional interactions (and in fact it must if those relationships
exist in the subsequent analysis models). In addition to parametric
models like the linear regression represented here, other imputation
approaches can be employed, including non-parametric methods such as
predictive mean matching, classification and regression trees (CART),
random forests, or k-nearest neighbors (Van Buuren and Van Buuren 2012).

When fitting imputation models when using doubly robust methods with a
binary exposure, we recommend fitting the model separately within each
exposure group to ensure that the imputation model is congenial with the
outcome model that typically is fit separately by exposure status (or
with an interaction). For example, when imputing the continuous
confounder \(Z_C\) separately within each exposure group \(X = 1\) and
\(X = 0\), we would specify two distinct models:

\begin{align*}
\textrm{For } X = 1:\quad  Z_C & = \beta_0^{(1)} + \beta_1^{(1)} Z_I + \beta_2^{(1)} Z_P + \beta_3^{(1)} Y + \varepsilon^{(1)}, \quad \varepsilon^{(1)} \sim \mathcal{N}(0, 1), \\
\textrm{For } X = 0:\quad  Z_C &  = \beta_0^{(0)} + \beta_1^{(0)} Z_I + \beta_2^{(0)} Z_P + \beta_3^{(0)} Y + \varepsilon^{(0)}, \quad \varepsilon^{(0)} \sim \mathcal{N}(0, 1).
\end{align*}

Each set of coefficients \(\boldsymbol\beta^{(x)}\) is estimated using
only data from units with exposure status \(X = x\). This stratified
approach ensures that the imputation model remains congenial with
outcome models that are fit separately by exposure group or that include
interactions with \(X\).

These imputation models are used to generate \(m\) completed datasets by
drawing from the posterior predictive distribution of the missing
values. Within each imputed dataset, causal estimands can be computed
using IPW alone or doubly robust methods described above
(Equation~\ref{eq-ipw} and Equation~\ref{eq-dr}). Estimates are then
combined using Rubin's rules (Rubin 1987). The pooled point estimate is:

\begin{equation}\phantomsection\label{eq-dr-mi}{
\bar{\Delta} = \frac{1}{m} \sum_{j=1}^m \hat{\Delta}^{(j)},
}\end{equation}

with total variance: \[
T = \bar{U} + \left(1 + \frac{1}{m} \right) B,
\]

where \(\bar{U} = \frac{1}{m} \sum_{j=1}^m U^{(j)}\) is the average
within-imputation variance, and
\(B = \frac{1}{m - 1} \sum_{j=1}^m \left( \hat{\Delta}^{(j)} - \bar{\Delta} \right)^2\)
is the between-imputation variance.

\section{Theoretical Considerations}\label{theoretical-considerations}

When implementing multiple imputation, ensuring compatibility between
the imputation and analysis models is essential for obtaining unbiased
estimates. In causal inference, one might assume that accurately
modeling the relationships among the exposure, outcome, and confounders
is sufficient, since these are the only variables are required to
identify the causal effect. However, this assumption can break down when
stochastic imputation (like MI) is used. Variables that are marginally
independent in the data-generating process can become conditionally
dependent once we condition on their common effect. If such variables
are omitted from the imputation model, it can lead to biased estimates,
even when the analysis models are correctly specified. We need to
carefully consider the full joint distribution of variables in the
analysis, not just those required for identification.

Consider our data generating mechanism (Figure~\ref{fig-dag}) where
\(X\) causes \(Y\), \(\mathbf{Z}_C\) are confounders affecting both
\(X\) and \(Y\) and \(\mathbf{Z}_P\) are precision variables that only
affects \(Y\). While \(\mathbf{Z}_C\) and \(\mathbf{Z}_P\) are
marginally independent, they will become conditionally associated if
conditioned on \(Y\) because they are both causes of \(Y\).

To motivate our recommendations for specifying imputation models in
causal inference with missing data, we consider two scenarios: (i)
missing confounders \(\mathbf{Z}_C\) and (ii) missing outcome \(Y\). In
all cases, we assume the missingness mechanism satisfies the MAR
assumption. We estimate the propensity score model as
\(\pi(\mathbf{Z}_C)\). The outcome component of the doubly robust
estimator is estimated as \({\mu}_0(\mathbf{{Z}_{{C}}}, \mathbf{Z}_P)\)
and \({\mu}_1(\mathbf{Z}_C, \mathbf{Z}_P)\) as in Equation~\ref{eq-dr},
and combined across imputed datasets as described in
Equation~\ref{eq-dr-mi}.

\subsection{Imputing Missing Confounders}\label{sec-imp-conf}

Suppose we have missing values in one of the confounders, denoted
\({Z}_{{C}(j)}\) indicating that the \(j\)-th column of \(\mathbf{Z}_C\)
(the matrix of all confounders) has missing entries. When handling
missing values in a confounder such as \(Z_{C(j)}\), the imputation
model must include all variables present in the analysis models to avoid
bias. This fundamental principle extends beyond merely including obvious
causal components (\(X\) and \(Y\)); it necessitates including all
auxiliary variables, such as precision variables \(\mathbf{Z}_P\), that
appear in the subsequent models. The rationale stems from the need to
preserve the correct correlation structure between the imputed and
observed variables. When imputing \(Z_{C(j)}\), including \(Y\) in the
imputation model is essential for unbiased estimation (D'Agostino
McGowan, Lotspeich, and Hepler 2024). However, this inclusion induces a
conditional association between variables that may be marginally
independent in the data-generating process (e.g., between \(Z_{C(j)}\)
and \(\mathbf{Z}_P\), see Figure~\ref{fig-dag}).

Doubly robust estimators provide theoretical protection against model
misspecification, requiring only one of two models, either the
propensity score model or the outcome model, to be correctly specified
for consistent estimation. However, this double robustness property does
not extend to protecting against imputation misspecification, even for
variables that appear exclusively in the outcome model and not in the
propensity score model.

Examining Equation~\ref{eq-dr} reveals that it comprises two components:
the estimated causal effect derived from the outcome model (first term)
and the weighted residuals from the outcome models (the weighted
difference between observed outcomes and their model-predicted values,
the second term). When precision variables (\(\mathbf{Z}_P\)) are
omitted from the imputation model but included in the outcome model, the
propensity score model itself remains correctly specified and unaffected
by \(\mathbf{Z}_P\). However, within strata defined by exposure and
confounders, the residuals from the outcome model will average to zero
(a fundamental property of least squares estimation). This mathematical
inevitability nullifies the second term of Equation~\ref{eq-dr},
effectively preventing the correctly specified propensity score model
from contributing to the estimate at all. The estimation then relies
solely on the outcome model's contribution, which yields biased
estimates because it uses incorrectly imputed data.

The proper imputation model for \(Z_{C(j)}\) must therefore reflect its
conditional distribution given all variables in subsequent analysis
models: the exposure (\(X\)), the outcome (\(Y\)), other confounders
(\(\mathbf{Z}_{C(j)}\)), and precision variables (\(\mathbf{Z}_P\)).

Since the doubly robust estimator in Equation~\ref{eq-dr} involves
fitting separate outcome models for exposed and unexposed groups, we can
separately fit the imputation model among the exposed and unexposed to
account for this.

\subsection{Imputing Missing Outcomes}\label{sec-imp-out}

As above, when imputing \(Y\) it is intuitive that \(X\) is necessary in
the imputation model. Likewise, if an estimator such as
Equation~\ref{eq-dr} is used that has two outcome models depending on
exposure status, the imputation models should also be fit separately
among exposure groups. Interestingly, when \(Y\) is the variable with
missing values requiring imputation, excluding \(\mathbf{Z}_P\) from the
imputation model does not introduce bias in estimating the \(X\)-\(Y\)
relationship, even when \(\mathbf{Z}_P\) is included in the analysis
model as long as separate imputation models are appropriately fit by
exposure status. Likewise, excluding \(\mathbf{Z}_C\) also will not
induce bias as long as imputation models are fit separately by exposure
status.

This contrast highlights an important asymmetry in how misspecified
imputation models affect causal estimates. When imputing confounders,
failing to account for conditional associations induced by collider
structures leads to bias. However, when imputing outcomes, omitting
variables that are in the outcome model (like confounders or precision
variables) may affect efficiency but will not bias in the causal
parameter of interest when using a doubly robust estimator as long as
the imputation model is properly fit within exposure strata.

In the following section we empirically demonstrate these theoretical
considerations. In addition to demonstrating the bias introduced by
leaving necessary variables out of the imputation model, we also examine
the impact of including the right variables but with the incorrect
functional form.

\section{Simulation}\label{simulation}

We conducted a simulation studies to demonstrate the impact of (1)
leaving variables out of the imputation model, and (2) including the
right variables but with incorrect functional form. For each scenario we
examined four different sample sizes (100, 500, 1,000, and 2,000) and
ran 500 simulations for each. This number was chosen to ensure adequate
precision in our simulation estimates, even for the smallest sample
size.

\subsection{Primary Scenarios}\label{sec-sim-1}

\begin{figure}[H]

\centering{

\pandocbounded{\includegraphics[keepaspectratio]{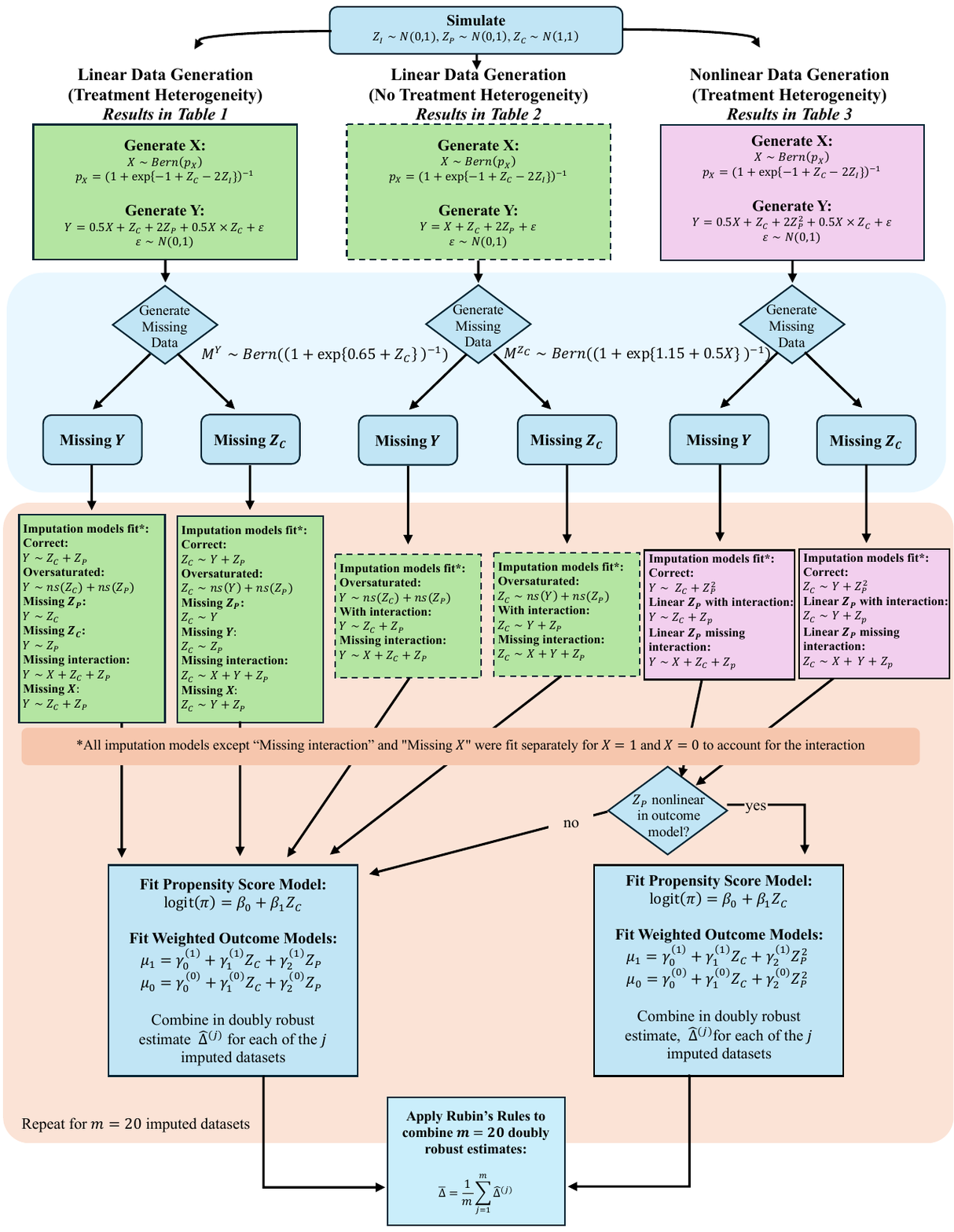}}

}

\caption{\label{fig-schema}Simulation Schema}

\end{figure}

Figure~\ref{fig-schema} presents the overall simulation schema. For each
scenario, we first generated three normally distributed variables
\(Z_I \sim N(0,1)\), \(Z_P \sim N(0,1)\), and \(Z_C \sim N(1,1)\). The
binary treatment variable \(X\) was generated from a Bernoulli
distribution with probability:

\[
p_X = \left(1 + \exp\{Z_C - 2Z_I\}\right)^{-1}.
\]

The outcome \(Y\) was generated according to:

\begin{align*}
\text{Linear case with treatment heterogeneity: } & Y = 0.5X + Z_C + 2Z_P + 0.5X\times Z_C + \varepsilon,\\
\text{Linear case without treatment heterogeneity: } & Y = X + Z_C + 2Z_P + \varepsilon,\\
\text{Nonlinear case with treatment heterogeneity: } & Y = 0.5X + Z_C + 2Z_P^2 + 0.5X\times Z_C +\varepsilon.
\end{align*}

where \(\varepsilon \sim N(0,1)\). To simulate missing data mechanisms,
we introduced missingness for \(Y\) and \(Z_C\) with the following
probabilities:

\begin{align*}
\mathbb{P}(M^Y = 1) &= \left(1 + \exp\{0.65 + Z_C\}\right)^{-1},\\
\mathbb{P}(M^{Z_C} = 1) &= \left(1 + \exp\{1.15+0.5X\}\right)^{-1}.
\end{align*}

where \(M^V = 1\) indicates that variable \(V\) is missing. The above
values were chosen to create approximately 20\% missingness in each
simulation.

Our propensity score model is specified as:

\begin{align*}
\text{logit}(\pi) = \beta_0 + \beta_1 Z_C.
\end{align*}

Outcome models are specified as:

\begin{align*}
\text{Linear case: } & \mu_1 = \gamma_0^{(1)} + \gamma_1^{(1)} Z_C + \gamma_2^{(1)} Z_P ,\\ 
&\mu_0 =  \gamma_0^{(0)} + \gamma_1^{(0)} Z_C + \gamma_2^{(0)} Z_P, \\
\text{Nonlinear case: } & \mu_1 = \gamma_0^{(1)} + \gamma_1^{(1)} Z_C + \gamma_2^{(1)} Z_P^2,\\
& \mu_0 = \gamma_0^{(0)} + \gamma_1^{(0)} Z_C + \gamma_2^{(0)} Z_P^2.
\end{align*}

For each missing data pattern, we determined and fitted the imputation
models as described in Figure~\ref{fig-schema} \(m=20\) times using the
mice package in R (van Buuren and Groothuis-Oudshoorn 2011). The
``oversaturated'' imputation model was split by exposure and included
\(Z_P\) and \(Z_C\), both fit with natural splines with 3 degrees of
freedom, when the outcome was missing and the outcome and \(Z_P\) fit
with a natural spline with three degrees of freedom when the confounder
was missing. We then calculated doubly robust estimates
\(\hat{\Delta}^{(j)}\) for each imputed dataset. These estimates were
combined using Rubin's rules to obtain the final estimate as described
in Section~\ref{sec-mi}. All data were generated such that the true
average treatment effect \(\Delta = 1\).

\newpage

\phantomsection\label{cell-fig-schema}

\subsection{Multiple Confounder
Scenarios}\label{multiple-confounder-scenarios}

\begin{figure}[H]

\centering{

\pandocbounded{\includegraphics[keepaspectratio]{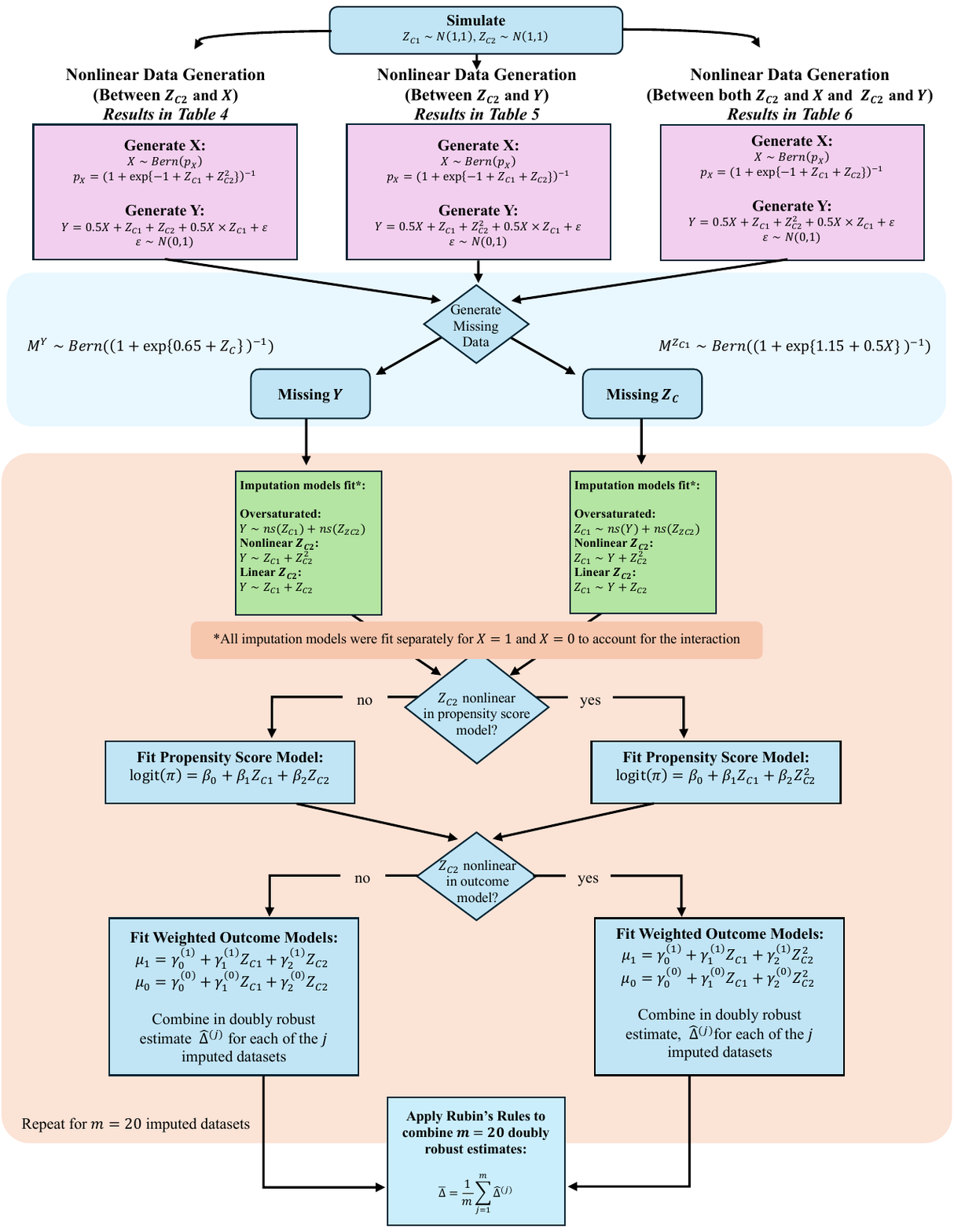}}

}

\caption{\label{fig-schema-2}Simulation Schema (Multiple Confounders)}

\end{figure}

Figure~\ref{fig-schema-2} presents the schema for our second simulation
study, which extends the framework to include two confounders with the
non-missing confounder having varying nonlinear relationships with the
exposure and/or outcome.

For each scenario, we first generated two normally distributed
variables: \(Z_{C1} \sim N(1,1)\) and \(Z_{C2} \sim N(1,1)\). We
examined three data generation scenarios that varied the nonlinear
relationships between \(Z_{C2}\) and the exposure and/or outcome.

\textbf{Scenario 1: Nonlinear \(Z_{C2}\)-\(X\) relationship}

The binary treatment variable \(X\) was generated from a Bernoulli
distribution with probability:

\begin{equation*}
p_X = \left(1 + \exp\{Z_{C1} + Z_{C2}^2\}\right)^{-1}.
\end{equation*}

The outcome \(Y\) was generated according to: \begin{equation*}
Y = 0.5X + Z_{C1} + Z_{C2} + 0.5X \times Z_{C1} + \varepsilon.
\end{equation*}

\textbf{Scenario 2: Nonlinear \(Z_{C2}\)-\(Y\) relationship}

The binary treatment variable \(X\) was generated from a Bernoulli
distribution with probability:

\begin{equation*}
p_X = \left(1 + \exp\{Z_{C1}+Z_{C2}\}\right)^{-1}.
\end{equation*}

The outcome \(Y\) was generated according to:

\begin{equation*}
Y = 0.5X + Z_{C1} + Z_{C2}^2 + 0.5X \times Z_{C1} + \varepsilon.
\end{equation*}

\textbf{Scenario 3: Nonlinear \(Z_{C2}\)-\(X\) and \(Z_{C2}\)-\(Y\)
relationships}

The binary treatment variable \(X\) was generated from a Bernoulli
distribution with probability:

\begin{equation*}
p_X = \left(1 + \exp\{Z_{C1} + Z_{C2}^2\}\right)^{-1}.
\end{equation*}

The outcome \(Y\) was generated according to:

\begin{equation*}
Y = 0.5X + Z_{C1} + Z_{C2}^2 + 0.5X \times Z_{C1} + \varepsilon,
\end{equation*}

where \(\varepsilon \sim N(0,1)\) in all cases.

To simulate missing data mechanisms, we introduced missingness for \(Y\)
and \(Z_{C1}\) using the same probabilities as in
Section~\ref{sec-sim-1}:

\begin{align*}
\mathbb{P}(M^Y = 1) &= \left(1 + \exp\{0.65 + Z_{C1}\}\right)^{-1},\\
\mathbb{P}(M^{Z_{C1}} = 1) &= \left(1 + \exp\{1.15+0.5X\}\right)^{-1},
\end{align*}

where \(M^V = 1\) indicates that variable \(V\) is missing. Again, the
above values were chosen to create approximately 20\% missingness in
each simulation.

Our propensity score models were correctly specified for each scenario
such that Scenarios 1 and 3 included the quadratic \(Z_{C2}\) term and
Scenario 2 included a linear \(Z_{C2}\) term.

Outcome models were similarly correctly specified such that Scenarios 2
and 3 included the quadratic \(Z_{C2}\) term and Scenario 1 included a
linear \(Z_{C2}\) term.

For each missing data pattern and scenario, we fit three different
imputation strategies, with separate models by treatment group \(X\) to
account for treatment heterogeneity: (1) an ``oversaturated'' imputation
model was split by exposure and included \(Z_{C1}\) and \(Z_{C2}\) ,
both fit with natural splines with 3 degrees of freedom, when the
outcome was missing and the outcome and \(Z_{C2}\) fit with a natural
spline with three degrees of freedom when \(Z_{C1}\) was missing, (2) an
imputation model with \(Z_{C2}\) included linearly and otherwise
correctly specified, and (3) an imputation model with \(Z_{C2}^2\)
included and otherwise correctly specified. Each imputation was
performed \(m=20\) times using the mice package in R, and doubly robust
estimates were combined using Rubin's rules to obtain the final estimate
as described in Section~\ref{sec-mi}. All data were generated such that
the true average treatment effect \(\Delta = 1\).

\phantomsection\label{cell-fig-schema-2}

\subsection{Results}\label{results}

\begin{figure}[H]

\centering{

\pandocbounded{\includegraphics[keepaspectratio]{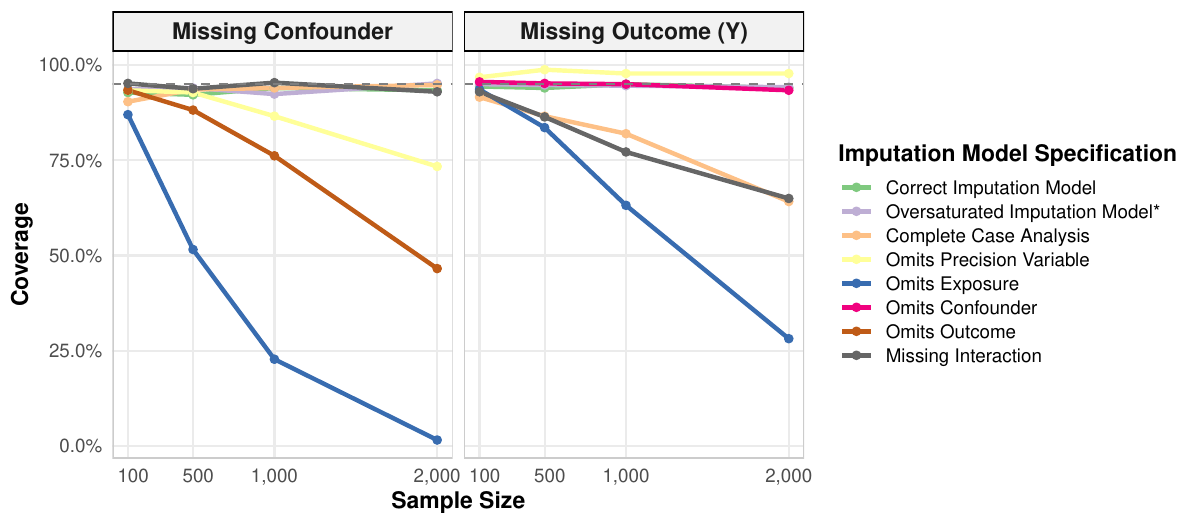}}

}

\caption{\label{fig-coverage}Coverage probability by sample size for
doubly robust estimates under different imputation specifications. Data
were simulated under the scenario described as ``Linear Data Generation
(Treatment Heterogenity)'' in Figure 2. The dashed line indicates
nominal 95\% coverage.}

\end{figure}

We conducted 500 simulation replications for each condition and computed
the mean and standard deviation of the estimated treatment effects.
Table~\ref{tbl-1} and Figure~\ref{fig-coverage} show the correctly
specified models as well as the impact of omitting \(X\), \(Y\),
\(Z_C\), and \(Z_P\) from the imputation models. Table~\ref{tbl-2} shows
the impact of the true data generating mechanism having homogeneous
treatment effects (instead of heterogeneous, as was the default).
Table~\ref{tbl-3} shows the impact of incorrectly specifying nonlinear
terms that were included in the analysis model in the imputation model.

\phantomsection\label{cell-fig-coverage}

The results reveal several important patterns in causal effect
estimation with missing data. Correctly specified imputation models
consistently yielded unbiased estimates of the average treatment effect
across all missingness scenarios, with mean estimates close to the true
value of \(\Delta = 1\) (Table~\ref{tbl-1}, Table~\ref{tbl-2},
Table~\ref{tbl-3}). Likewise, the oversaturated imputation model,
despite containing additional nonlinear terms, was also unbiased and had
proper coverage. Complete case analysis was unbiased when \(Z_C\) was
missing because the outcome model was correctly specified and included
\(X\), which was the factor that dictated whether \(Z_C\) was missing.
However, complete case analysis was biased when \(Y\) was missing
despite the fact that the missingness was at random (based on \(Z_C\)).
This underscores the importance of understanding the relationship
between the missing data mechanism and the causal structure when
deciding whether complete case analysis is appropriate.

When \(Z_C\) (confounder) was missing, omitting \(Z_P\) from the
imputation model resulted in substantial bias in the estimated effect
between \(X\) and \(Y\), despite \(Z_P\) being a precision variable that
affects only the outcome. Omitting \(Y\) or \(X\) when \(Z_C\) was
missing likewise produced biased estimate (Table~\ref{tbl-1},
Figure~\ref{fig-coverage}).

When \(Y\) (outcome) was missing, the patterns differed. Omitting
\(Z_C\) or \(Z_P\) had minimal impact on bias (as would be expected from
Section~\ref{sec-imp-out}), but omitting \(X\) biased estimates
downward. While omitting \(Z_P\) did not have an impact on the bias with
respect to the relationship between \(X\) and \(Y\), it did have a large
impact on the precision, as indicated by higher average standard
deviation (Table~\ref{tbl-1}).

Since we used a doubly robust estimator that effectively included an
interaction in the outcome model, for the imputation model to be
congenial, we needed this interaction in the imputation model too. One
approach to handle this is to separately fit the imputation model among
the exposed and unexposed groups, which is what we did here. This was
necessary for estimating the causal effect when \(Y\) was missing, but
not for the scenarios we examined when \(Z_C\) was missing. This
separate modeling approach was also not required when the true data
generating mechanism did not have treatment effect heterogeneity
(Table~\ref{tbl-2}).

In nonlinear scenarios, misspecifying the functional form of variables
produced similar bias to omitting them entirely. For instance, using the
wrong form for \(Z_P\) when \(Z_C\) is missing yielded bias of the same
magnitude as completely omitting \(Z_P\) (Table~\ref{tbl-3}). We also
examined a scenario where both the nonlinear term was misspecified as
linear and the interaction between \(X\) and \(Z_C\) was not properly
accounted for in the imputation model. Because omitting the nonlinearity
biases results downward while omitting the interaction biases results
upward, these opposing biases partially offset each other, resulting in
estimates that appear slightly better than those from misspecifying
nonlinearity alone (Table~\ref{tbl-3}). This phenomenon illustrates an
important methodological point that apparently ``good'' performance from
a misspecified model may actually reflect the fortuitous cancellation of
multiple sources of bias rather than correct model specification. Such
scenarios have led to misinterpretation of results in previous studies
(Mitra and Reiter 2016), where researchers may incorrectly conclude that
particular methods are adequate when the observed performance actually
stems from compensating errors.

Examining scenarios with multiple confounders where one was missing and
the other had various nonlinear relationships with the exposure and
outcome (Table~\ref{tbl-4}, Table~\ref{tbl-5}, Table~\ref{tbl-6}), we
find that imputation model sensitivity depends on whether the nonlinear
relationship involves the exposure or the outcome. When the nonlinear
relationship is between the additional confounder (\(Z_{C2}\)) and the
exposure, the imputation model is not sensitive to this misspecification
(assuming separate imputation models are correctly specified by exposure
status) (Table~\ref{tbl-4}). However, when the nonlinear relationship is
between the additional confounder and the outcome, incorrectly
specifying \(Z_{C2}\) in the imputation model as linear results in
substantial bias (Table~\ref{tbl-5} and Table~\ref{tbl-6}). Across all
scenarios, the oversaturated imputation model recovers unbiased
estimates provided the sample size is sufficient (i.e., it performs
poorly when n = 100).

Overall, these findings confirm that proper imputation model
specification requires including all variables that appear in either the
propensity score model or outcome model, in their correct functional
form. The bias patterns demonstrate that omitting seemingly secondary
variables like precision variables (\(Z_P\)) can substantially
compromise causal effect estimation as well as highlight the importance
of considering the functional form (the presence or interactions an
nonlinear terms, for example) in the subsequent models.

\begingroup
\setlength\LTleft{0\linewidth}
\setlength\LTright{0\linewidth}\fontsize{12.0pt}{14.4pt}\selectfont
\setlength{\LTpost}{0mm}

\begin{longtable}{@{\extracolsep{\fill}}>{\raggedright\arraybackslash}p{\dimexpr 0.30\linewidth -2\tabcolsep-1.5\arrayrulewidth}>{\raggedleft\arraybackslash}p{\dimexpr 0.12\linewidth -2\tabcolsep-1.5\arrayrulewidth}>{\raggedleft\arraybackslash}p{\dimexpr 0.12\linewidth -2\tabcolsep-1.5\arrayrulewidth}>{\raggedleft\arraybackslash}p{\dimexpr 0.12\linewidth -2\tabcolsep-1.5\arrayrulewidth}>{\raggedleft\arraybackslash}p{\dimexpr 0.12\linewidth -2\tabcolsep-1.5\arrayrulewidth}>{\raggedleft\arraybackslash}p{\dimexpr 0.12\linewidth -2\tabcolsep-1.5\arrayrulewidth}>{\raggedleft\arraybackslash}p{\dimexpr 0.12\linewidth -2\tabcolsep-1.5\arrayrulewidth}}

\caption{\label{tbl-1}Simulation results under heterogeneous treatment
effect data generation for the linear data generating mechanism by
imputation model specification. The true average treatment effect is 1.}

\tabularnewline

\toprule
Model Specification & Est. & MC SE & Avg. SE & Bias & RMSE & 95\% Cov. \\ 
\midrule 
\endfirsthead
\toprule
Model Specification & Est. & MC SE & Avg. SE & Bias & RMSE & 95\% Cov. \\ 
\midrule 
\endhead
\midrule\addlinespace[2.5pt]
\multicolumn{7}{>{\raggedright\arraybackslash}m{1\linewidth}}{Missing Confounder ($Z_C$) - n = 100} \\[2.5pt] 
\midrule\addlinespace[2.5pt]
Correct Imputation Model & 1.01 & 0.24 & 0.22 & 0.01 & 0.24 & 0.93 \\ 
Oversaturated Imputation Model* & 0.97 & 0.25 & 0.23 & -0.03 & 0.25 & 0.94 \\ 
Complete Case Analysis & 0.98 & 0.26 & 0.22 & -0.02 & 0.26 & 0.90 \\ 
Omits Precision Variable & 0.91 & 0.26 & 0.26 & -0.09 & 0.28 & 0.93 \\ 
Omits Exposure & 0.79 & 0.20 & 0.23 & -0.21 & 0.30 & 0.87 \\ 
Omits Outcome & 0.85 & 0.27 & 0.29 & -0.15 & 0.30 & 0.93 \\ 
Missing Interaction & 1.02 & 0.22 & 0.22 & 0.02 & 0.22 & 0.95 \\ 
\midrule\addlinespace[2.5pt]
\multicolumn{7}{>{\raggedright\arraybackslash}m{1\linewidth}}{Missing Confounder ($Z_C$) - n = 500} \\[2.5pt] 
\midrule\addlinespace[2.5pt]
Correct Imputation Model & 0.99 & 0.11 & 0.10 & -0.01 & 0.11 & 0.92 \\ 
Oversaturated Imputation Model* & 0.99 & 0.10 & 0.10 & -0.01 & 0.10 & 0.94 \\ 
Complete Case Analysis & 0.99 & 0.11 & 0.10 & -0.01 & 0.11 & 0.94 \\ 
Omits Precision Variable & 0.92 & 0.11 & 0.12 & -0.08 & 0.13 & 0.93 \\ 
Omits Exposure & 0.80 & 0.09 & 0.10 & -0.20 & 0.22 & 0.52 \\ 
Omits Outcome & 0.87 & 0.11 & 0.13 & -0.13 & 0.17 & 0.88 \\ 
Missing Interaction & 0.99 & 0.11 & 0.10 & -0.01 & 0.11 & 0.94 \\ 
\midrule\addlinespace[2.5pt]
\multicolumn{7}{>{\raggedright\arraybackslash}m{1\linewidth}}{Missing Confounder ($Z_C$) - n = 1000} \\[2.5pt] 
\midrule\addlinespace[2.5pt]
Correct Imputation Model & 1.00 & 0.07 & 0.07 & 0.00 & 0.07 & 0.94 \\ 
Oversaturated Imputation Model* & 1.00 & 0.08 & 0.07 & 0.00 & 0.08 & 0.92 \\ 
Complete Case Analysis & 0.98 & 0.07 & 0.07 & -0.02 & 0.08 & 0.94 \\ 
Omits Precision Variable & 0.92 & 0.08 & 0.09 & -0.08 & 0.11 & 0.87 \\ 
Omits Exposure & 0.80 & 0.06 & 0.07 & -0.20 & 0.21 & 0.23 \\ 
Omits Outcome & 0.87 & 0.08 & 0.09 & -0.13 & 0.16 & 0.76 \\ 
Missing Interaction & 1.00 & 0.07 & 0.07 & 0.00 & 0.07 & 0.95 \\ 
\midrule\addlinespace[2.5pt]
\multicolumn{7}{>{\raggedright\arraybackslash}m{1\linewidth}}{Missing Confounder ($Z_C$) - n = 2000} \\[2.5pt] 
\midrule\addlinespace[2.5pt]
Correct Imputation Model & 0.99 & 0.05 & 0.05 & -0.01 & 0.05 & 0.93 \\ 
Oversaturated Imputation Model* & 1.00 & 0.05 & 0.05 & 0.00 & 0.05 & 0.95 \\ 
Complete Case Analysis & 0.99 & 0.05 & 0.05 & -0.01 & 0.05 & 0.95 \\ 
Omits Precision Variable & 0.92 & 0.05 & 0.06 & -0.08 & 0.10 & 0.73 \\ 
Omits Exposure & 0.80 & 0.05 & 0.05 & -0.20 & 0.21 & 0.02 \\ 
Omits Outcome & 0.86 & 0.05 & 0.07 & -0.14 & 0.15 & 0.47 \\ 
Missing Interaction & 1.00 & 0.05 & 0.05 & 0.00 & 0.05 & 0.93 \\ 
\midrule\addlinespace[2.5pt]
\multicolumn{7}{>{\raggedright\arraybackslash}m{1\linewidth}}{Missing Outcome (Y) - n = 100} \\[2.5pt] 
\midrule\addlinespace[2.5pt]
Correct Imputation Model & 1.02 & 0.24 & 0.24 & 0.02 & 0.24 & 0.94 \\ 
Oversaturated Imputation Model* & 0.99 & 0.27 & 0.29 & -0.01 & 0.27 & 0.95 \\ 
Complete Case Analysis & 1.09 & 0.25 & 0.22 & 0.09 & 0.26 & 0.92 \\ 
Omits Precision Variable & 0.99 & 0.36 & 0.42 & -0.01 & 0.36 & 0.97 \\ 
Omits Exposure & 0.84 & 0.20 & 0.24 & -0.16 & 0.26 & 0.94 \\ 
Omits Confounder & 0.96 & 0.28 & 0.31 & -0.04 & 0.28 & 0.96 \\ 
Missing Interaction & 1.07 & 0.24 & 0.24 & 0.07 & 0.25 & 0.93 \\ 
\midrule\addlinespace[2.5pt]
\multicolumn{7}{>{\raggedright\arraybackslash}m{1\linewidth}}{Missing Outcome (Y) - n = 500} \\[2.5pt] 
\midrule\addlinespace[2.5pt]
Correct Imputation Model & 1.00 & 0.11 & 0.11 & 0.00 & 0.11 & 0.94 \\ 
Oversaturated Imputation Model* & 0.99 & 0.11 & 0.11 & -0.01 & 0.11 & 0.95 \\ 
Complete Case Analysis & 1.08 & 0.11 & 0.10 & 0.08 & 0.13 & 0.87 \\ 
Omits Precision Variable & 1.00 & 0.15 & 0.19 & 0.00 & 0.15 & 0.99 \\ 
Omits Exposure & 0.87 & 0.09 & 0.11 & -0.13 & 0.16 & 0.84 \\ 
Omits Confounder & 0.96 & 0.13 & 0.14 & -0.04 & 0.13 & 0.95 \\ 
Missing Interaction & 1.08 & 0.11 & 0.11 & 0.08 & 0.14 & 0.86 \\ 
\midrule\addlinespace[2.5pt]
\multicolumn{7}{>{\raggedright\arraybackslash}m{1\linewidth}}{Missing Outcome (Y) - n = 1000} \\[2.5pt] 
\midrule\addlinespace[2.5pt]
Correct Imputation Model & 0.99 & 0.08 & 0.08 & -0.01 & 0.08 & 0.95 \\ 
Oversaturated Imputation Model* & 1.00 & 0.08 & 0.08 & 0.00 & 0.08 & 0.95 \\ 
Complete Case Analysis & 1.08 & 0.07 & 0.07 & 0.08 & 0.11 & 0.82 \\ 
Omits Precision Variable & 1.00 & 0.11 & 0.13 & 0.00 & 0.11 & 0.98 \\ 
Omits Exposure & 0.87 & 0.06 & 0.08 & -0.13 & 0.15 & 0.63 \\ 
Omits Confounder & 0.96 & 0.09 & 0.10 & -0.04 & 0.10 & 0.95 \\ 
Missing Interaction & 1.09 & 0.07 & 0.08 & 0.09 & 0.11 & 0.77 \\ 
\midrule\addlinespace[2.5pt]
\multicolumn{7}{>{\raggedright\arraybackslash}m{1\linewidth}}{Missing Outcome (Y) - n = 2000} \\[2.5pt] 
\midrule\addlinespace[2.5pt]
Correct Imputation Model & 1.00 & 0.05 & 0.05 & 0.00 & 0.05 & 0.94 \\ 
Oversaturated Imputation Model* & 1.00 & 0.06 & 0.05 & 0.00 & 0.06 & 0.94 \\ 
Complete Case Analysis & 1.08 & 0.05 & 0.05 & 0.08 & 0.10 & 0.64 \\ 
Omits Precision Variable & 1.00 & 0.08 & 0.09 & 0.00 & 0.08 & 0.98 \\ 
Omits Exposure & 0.87 & 0.04 & 0.06 & -0.13 & 0.14 & 0.28 \\ 
Omits Confounder & 0.96 & 0.06 & 0.07 & -0.04 & 0.08 & 0.93 \\ 
Missing Interaction & 1.08 & 0.05 & 0.05 & 0.08 & 0.10 & 0.65 \\ 
\bottomrule

\end{longtable}

\begin{minipage}{\linewidth}
Abbreviations: n = Sample size; Est. = Estimate; MC SE = Monte Carlo Standard Error; Avg. SE = Average Estimated Standard Error; 95\% Cov. = 95\% Confidence Interval Coverage. *Oversaturated Imputation Models fit split by exposure, include the outcome, and include the precision and confounder fit with natural splines with three degrees of freedom each.\\
\end{minipage}
\endgroup

\begingroup
\setlength\LTleft{0\linewidth}
\setlength\LTright{0\linewidth}\fontsize{12.0pt}{14.4pt}\selectfont
\setlength{\LTpost}{0mm}

\begin{longtable}{@{\extracolsep{\fill}}>{\raggedright\arraybackslash}p{\dimexpr 0.30\linewidth -2\tabcolsep-1.5\arrayrulewidth}>{\raggedleft\arraybackslash}p{\dimexpr 0.12\linewidth -2\tabcolsep-1.5\arrayrulewidth}>{\raggedleft\arraybackslash}p{\dimexpr 0.12\linewidth -2\tabcolsep-1.5\arrayrulewidth}>{\raggedleft\arraybackslash}p{\dimexpr 0.12\linewidth -2\tabcolsep-1.5\arrayrulewidth}>{\raggedleft\arraybackslash}p{\dimexpr 0.12\linewidth -2\tabcolsep-1.5\arrayrulewidth}>{\raggedleft\arraybackslash}p{\dimexpr 0.12\linewidth -2\tabcolsep-1.5\arrayrulewidth}>{\raggedleft\arraybackslash}p{\dimexpr 0.12\linewidth -2\tabcolsep-1.5\arrayrulewidth}}

\caption{\label{tbl-2}Simulation results under homogenous treatment
effect data generation by imputation model specification. The true
average treatment effect is 1.}

\tabularnewline

\toprule
Model Specification & Est. & MC SE & Avg. SE & Bias & RMSE & 95\% Cov. \\ 
\midrule 
\endfirsthead
\toprule
Model Specification & Est. & MC SE & Avg. SE & Bias & RMSE & 95\% Cov. \\ 
\midrule 
\endhead
\midrule\addlinespace[2.5pt]
\multicolumn{7}{>{\raggedright\arraybackslash}m{1\linewidth}}{Missing Confounder ($Z_C$) - n = 100} \\[2.5pt] 
\midrule\addlinespace[2.5pt]
Correct Imputation Model & 1.01 & 0.23 & 0.22 & 0.01 & 0.23 & 0.93 \\ 
Oversaturated Imputation Model* & 0.96 & 0.24 & 0.23 & -0.04 & 0.24 & 0.94 \\ 
Complete Case Analysis & 0.99 & 0.25 & 0.22 & -0.01 & 0.25 & 0.91 \\ 
Missing Interaction & 1.02 & 0.22 & 0.22 & 0.02 & 0.22 & 0.95 \\ 
\midrule\addlinespace[2.5pt]
\multicolumn{7}{>{\raggedright\arraybackslash}m{1\linewidth}}{Missing Confounder ($Z_C$) - n = 500} \\[2.5pt] 
\midrule\addlinespace[2.5pt]
Correct Imputation Model & 1.00 & 0.10 & 0.10 & 0.00 & 0.10 & 0.94 \\ 
Oversaturated Imputation Model* & 0.99 & 0.10 & 0.10 & -0.01 & 0.10 & 0.95 \\ 
Complete Case Analysis & 0.99 & 0.11 & 0.10 & -0.01 & 0.11 & 0.94 \\ 
Missing Interaction & 0.99 & 0.10 & 0.10 & -0.01 & 0.10 & 0.94 \\ 
\midrule\addlinespace[2.5pt]
\multicolumn{7}{>{\raggedright\arraybackslash}m{1\linewidth}}{Missing Confounder ($Z_C$) - n = 1000} \\[2.5pt] 
\midrule\addlinespace[2.5pt]
Correct Imputation Model & 1.00 & 0.07 & 0.07 & 0.00 & 0.07 & 0.94 \\ 
Oversaturated Imputation Model* & 1.00 & 0.07 & 0.07 & 0.00 & 0.07 & 0.94 \\ 
Complete Case Analysis & 0.99 & 0.07 & 0.07 & -0.01 & 0.07 & 0.95 \\ 
Missing Interaction & 1.00 & 0.07 & 0.07 & 0.00 & 0.07 & 0.97 \\ 
\midrule\addlinespace[2.5pt]
\multicolumn{7}{>{\raggedright\arraybackslash}m{1\linewidth}}{Missing Confounder ($Z_C$) - n = 2000} \\[2.5pt] 
\midrule\addlinespace[2.5pt]
Correct Imputation Model & 0.99 & 0.05 & 0.05 & -0.01 & 0.05 & 0.94 \\ 
Oversaturated Imputation Model* & 1.00 & 0.05 & 0.05 & 0.00 & 0.05 & 0.95 \\ 
Complete Case Analysis & 1.00 & 0.05 & 0.05 & 0.00 & 0.05 & 0.96 \\ 
Missing Interaction & 1.00 & 0.05 & 0.05 & 0.00 & 0.05 & 0.95 \\ 
\midrule\addlinespace[2.5pt]
\multicolumn{7}{>{\raggedright\arraybackslash}m{1\linewidth}}{Missing Outcome (Y) - n = 100} \\[2.5pt] 
\midrule\addlinespace[2.5pt]
Correct Imputation Model & 1.01 & 0.24 & 0.24 & 0.01 & 0.24 & 0.95 \\ 
Oversaturated Imputation Model* & 0.99 & 0.26 & 0.29 & -0.01 & 0.26 & 0.97 \\ 
Complete Case Analysis & 1.00 & 0.24 & 0.22 & 0.00 & 0.24 & 0.94 \\ 
Missing Interaction & 0.99 & 0.24 & 0.23 & -0.01 & 0.24 & 0.95 \\ 
\midrule\addlinespace[2.5pt]
\multicolumn{7}{>{\raggedright\arraybackslash}m{1\linewidth}}{Missing Outcome (Y) - n = 500} \\[2.5pt] 
\midrule\addlinespace[2.5pt]
Correct Imputation Model & 1.00 & 0.11 & 0.11 & 0.00 & 0.11 & 0.95 \\ 
Oversaturated Imputation Model* & 0.99 & 0.10 & 0.11 & -0.01 & 0.10 & 0.96 \\ 
Complete Case Analysis & 0.99 & 0.11 & 0.10 & -0.01 & 0.11 & 0.94 \\ 
Missing Interaction & 1.00 & 0.11 & 0.11 & 0.00 & 0.11 & 0.94 \\ 
\midrule\addlinespace[2.5pt]
\multicolumn{7}{>{\raggedright\arraybackslash}m{1\linewidth}}{Missing Outcome (Y) - n = 1000} \\[2.5pt] 
\midrule\addlinespace[2.5pt]
Correct Imputation Model & 0.99 & 0.08 & 0.08 & -0.01 & 0.08 & 0.96 \\ 
Oversaturated Imputation Model* & 1.00 & 0.08 & 0.08 & 0.00 & 0.08 & 0.96 \\ 
Complete Case Analysis & 0.99 & 0.07 & 0.07 & -0.01 & 0.07 & 0.95 \\ 
Missing Interaction & 1.00 & 0.07 & 0.07 & 0.00 & 0.07 & 0.96 \\ 
\midrule\addlinespace[2.5pt]
\multicolumn{7}{>{\raggedright\arraybackslash}m{1\linewidth}}{Missing Outcome (Y) - n = 2000} \\[2.5pt] 
\midrule\addlinespace[2.5pt]
Correct Imputation Model & 1.00 & 0.05 & 0.05 & 0.00 & 0.05 & 0.95 \\ 
Oversaturated Imputation Model* & 1.00 & 0.05 & 0.05 & 0.00 & 0.05 & 0.94 \\ 
Complete Case Analysis & 1.00 & 0.05 & 0.05 & 0.00 & 0.05 & 0.95 \\ 
Missing Interaction & 1.00 & 0.05 & 0.05 & 0.00 & 0.05 & 0.95 \\ 
\bottomrule

\end{longtable}

\begin{minipage}{\linewidth}
Abbreviations: n = Sample size; Est. = Estimate; MC SE = Monte Carlo Standard Error; Avg. SE = Average Estimated Standard Error; 95\% Cov. = 95\% Confidence Interval Coverage. *Oversaturated Imputation Models fit split by exposure, include the outcome, and include the precision and confounder fit with natural splines with three degrees of freedom each.\\
\end{minipage}
\endgroup

\begingroup
\setlength\LTleft{0\linewidth}
\setlength\LTright{0\linewidth}\fontsize{12.0pt}{14.4pt}\selectfont
\setlength{\LTpost}{0mm}

\begin{longtable}{@{\extracolsep{\fill}}>{\raggedright\arraybackslash}p{\dimexpr 0.30\linewidth -2\tabcolsep-1.5\arrayrulewidth}>{\raggedleft\arraybackslash}p{\dimexpr 0.12\linewidth -2\tabcolsep-1.5\arrayrulewidth}>{\raggedleft\arraybackslash}p{\dimexpr 0.12\linewidth -2\tabcolsep-1.5\arrayrulewidth}>{\raggedleft\arraybackslash}p{\dimexpr 0.12\linewidth -2\tabcolsep-1.5\arrayrulewidth}>{\raggedleft\arraybackslash}p{\dimexpr 0.12\linewidth -2\tabcolsep-1.5\arrayrulewidth}>{\raggedleft\arraybackslash}p{\dimexpr 0.12\linewidth -2\tabcolsep-1.5\arrayrulewidth}>{\raggedleft\arraybackslash}p{\dimexpr 0.12\linewidth -2\tabcolsep-1.5\arrayrulewidth}}

\caption{\label{tbl-3}Simulation results when the data were generated
with a nonlinear relationship between \(Z_P\) and \(Y\) by imputation
model specification. The true average treatment effect is 1.}

\tabularnewline

\toprule
Model Specification & Est. & MC SE & Avg. SE & Bias & RMSE & 95\% Cov. \\ 
\midrule 
\endfirsthead
\toprule
Model Specification & Est. & MC SE & Avg. SE & Bias & RMSE & 95\% Cov. \\ 
\midrule 
\endhead
\midrule\addlinespace[2.5pt]
\multicolumn{7}{>{\raggedright\arraybackslash}m{1\linewidth}}{Missing Confounder ($Z_C$) - n = 100} \\[2.5pt] 
\midrule\addlinespace[2.5pt]
Correct Nonlinear Imputation Model & 1.02 & 0.45 & 0.39 & 0.02 & 0.45 & 0.91 \\ 
Misspecified Precision Variable & 0.85 & 0.54 & 0.49 & -0.15 & 0.56 & 0.91 \\ 
Misspecified Precision Variable, Missing Interaction & 0.92 & 0.46 & 0.46 & -0.08 & 0.46 & 0.94 \\ 
Precision Variable Linear in both Imputatation and Analysis Model & 1.03 & 0.71 & 0.60 & 0.03 & 0.71 & 0.92 \\ 
\midrule\addlinespace[2.5pt]
\multicolumn{7}{>{\raggedright\arraybackslash}m{1\linewidth}}{Missing Confounder ($Z_C$) - n = 500} \\[2.5pt] 
\midrule\addlinespace[2.5pt]
Correct Nonlinear Imputation Model & 0.98 & 0.19 & 0.17 & -0.02 & 0.19 & 0.91 \\ 
Misspecified Precision Variable & 0.80 & 0.24 & 0.24 & -0.20 & 0.32 & 0.88 \\ 
Misspecified Precision Variable, Missing Interaction & 0.86 & 0.21 & 0.23 & -0.14 & 0.25 & 0.93 \\ 
Precision Variable Linear in both Imputatation and Analysis Model & 0.99 & 0.29 & 0.29 & -0.01 & 0.29 & 0.95 \\ 
\midrule\addlinespace[2.5pt]
\multicolumn{7}{>{\raggedright\arraybackslash}m{1\linewidth}}{Missing Confounder ($Z_C$) - n = 1000} \\[2.5pt] 
\midrule\addlinespace[2.5pt]
Correct Nonlinear Imputation Model & 1.01 & 0.13 & 0.12 & 0.01 & 0.13 & 0.95 \\ 
Misspecified Precision Variable & 0.81 & 0.17 & 0.18 & -0.19 & 0.25 & 0.87 \\ 
Misspecified Precision Variable, Missing Interaction & 0.85 & 0.16 & 0.17 & -0.15 & 0.22 & 0.91 \\ 
Precision Variable Linear in both Imputatation and Analysis Model & 1.01 & 0.21 & 0.21 & 0.01 & 0.21 & 0.95 \\ 
\midrule\addlinespace[2.5pt]
\multicolumn{7}{>{\raggedright\arraybackslash}m{1\linewidth}}{Missing Confounder ($Z_C$) - n = 2000} \\[2.5pt] 
\midrule\addlinespace[2.5pt]
Correct Nonlinear Imputation Model & 1.00 & 0.09 & 0.09 & 0.00 & 0.09 & 0.94 \\ 
Misspecified Precision Variable & 0.81 & 0.12 & 0.13 & -0.19 & 0.22 & 0.76 \\ 
Misspecified Precision Variable, Missing Interaction & 0.85 & 0.12 & 0.13 & -0.15 & 0.19 & 0.85 \\ 
Precision Variable Linear in both Imputatation and Analysis Model & 0.99 & 0.14 & 0.15 & -0.01 & 0.14 & 0.95 \\ 
\midrule\addlinespace[2.5pt]
\multicolumn{7}{>{\raggedright\arraybackslash}m{1\linewidth}}{Missing Outcome (Y) - n = 100} \\[2.5pt] 
\midrule\addlinespace[2.5pt]
Correct Nonlinear Imputation Model & 1.01 & 0.48 & 0.42 & 0.01 & 0.48 & 0.92 \\ 
Misspecified Precision Variable & 0.89 & 1.42 & 1.10 & -0.11 & 1.42 & 0.93 \\ 
Misspecified Precision Variable, Missing Interaction & 0.88 & 1.49 & 1.09 & -0.12 & 1.49 & 0.94 \\ 
\midrule\addlinespace[2.5pt]
\multicolumn{7}{>{\raggedright\arraybackslash}m{1\linewidth}}{Missing Outcome (Y) - n = 500} \\[2.5pt] 
\midrule\addlinespace[2.5pt]
Correct Nonlinear Imputation Model & 1.00 & 0.18 & 0.18 & 0.00 & 0.18 & 0.94 \\ 
Misspecified Precision Variable & 0.91 & 0.66 & 0.62 & -0.09 & 0.67 & 0.95 \\ 
Misspecified Precision Variable, Missing Interaction & 1.01 & 0.67 & 0.63 & 0.01 & 0.67 & 0.96 \\ 
\midrule\addlinespace[2.5pt]
\multicolumn{7}{>{\raggedright\arraybackslash}m{1\linewidth}}{Missing Outcome (Y) - n = 1000} \\[2.5pt] 
\midrule\addlinespace[2.5pt]
Correct Nonlinear Imputation Model & 1.00 & 0.13 & 0.13 & 0.00 & 0.13 & 0.94 \\ 
Misspecified Precision Variable & 0.96 & 0.49 & 0.47 & -0.04 & 0.49 & 0.95 \\ 
Misspecified Precision Variable, Missing Interaction & 1.07 & 0.50 & 0.49 & 0.07 & 0.50 & 0.97 \\ 
\midrule\addlinespace[2.5pt]
\multicolumn{7}{>{\raggedright\arraybackslash}m{1\linewidth}}{Missing Outcome (Y) - n = 2000} \\[2.5pt] 
\midrule\addlinespace[2.5pt]
Correct Nonlinear Imputation Model & 1.00 & 0.09 & 0.09 & 0.00 & 0.09 & 0.96 \\ 
Misspecified Precision Variable & 1.01 & 0.36 & 0.36 & 0.01 & 0.36 & 0.96 \\ 
Misspecified Precision Variable, Missing Interaction & 1.05 & 0.33 & 0.35 & 0.05 & 0.33 & 0.97 \\ 
\bottomrule

\end{longtable}

\begin{minipage}{\linewidth}
Abbreviations: n = Sample size; Est. = Estimate; MC SE = Monte Carlo Standard Error; Avg. SE = Average Estimated Standard Error; 95\% Cov. = 95\% Confidence Interval Coverage.\\
\end{minipage}
\endgroup

\begingroup
\setlength\LTleft{0\linewidth}
\setlength\LTright{0\linewidth}\fontsize{12.0pt}{14.4pt}\selectfont
\setlength{\LTpost}{0mm}

\begin{longtable}{@{\extracolsep{\fill}}>{\raggedright\arraybackslash}p{\dimexpr 0.30\linewidth -2\tabcolsep-1.5\arrayrulewidth}>{\raggedleft\arraybackslash}p{\dimexpr 0.12\linewidth -2\tabcolsep-1.5\arrayrulewidth}>{\raggedleft\arraybackslash}p{\dimexpr 0.12\linewidth -2\tabcolsep-1.5\arrayrulewidth}>{\raggedleft\arraybackslash}p{\dimexpr 0.12\linewidth -2\tabcolsep-1.5\arrayrulewidth}>{\raggedleft\arraybackslash}p{\dimexpr 0.12\linewidth -2\tabcolsep-1.5\arrayrulewidth}>{\raggedleft\arraybackslash}p{\dimexpr 0.12\linewidth -2\tabcolsep-1.5\arrayrulewidth}>{\raggedleft\arraybackslash}p{\dimexpr 0.12\linewidth -2\tabcolsep-1.5\arrayrulewidth}}

\caption{\label{tbl-4}Simulation results when the data were generated
with a nonlinear relationship between \(Z_{C2}\) and \(X\) by imputation
model specification. The true average treatment effect is 1.}

\tabularnewline

\toprule
Model Specification & Est. & MC SE & Avg. SE & Bias & RMSE & 95\% Cov. \\ 
\midrule 
\endfirsthead
\toprule
Model Specification & Est. & MC SE & Avg. SE & Bias & RMSE & 95\% Cov. \\ 
\midrule 
\endhead
\midrule\addlinespace[2.5pt]
\multicolumn{7}{>{\raggedright\arraybackslash}m{1\linewidth}}{Missing Confounder ($Z_{C1}$) - n = 100} \\[2.5pt] 
\midrule\addlinespace[2.5pt]
Correct Imputation Model & 0.97 & 0.40 & 0.33 & -0.03 & 0.40 & 0.87 \\ 
Oversaturated Imputation Model* & 0.92 & 0.44 & 0.34 & -0.08 & 0.45 & 0.82 \\ 
Misspecified Confounder ($Z_{C2}$) & 1.01 & 0.40 & 0.31 & 0.01 & 0.40 & 0.85 \\ 
\midrule\addlinespace[2.5pt]
\multicolumn{7}{>{\raggedright\arraybackslash}m{1\linewidth}}{Missing Confounder ($Z_{C1}$) - n = 500} \\[2.5pt] 
\midrule\addlinespace[2.5pt]
Correct Imputation Model & 0.97 & 0.21 & 0.18 & -0.03 & 0.21 & 0.87 \\ 
Oversaturated Imputation Model* & 0.98 & 0.21 & 0.17 & -0.02 & 0.21 & 0.86 \\ 
Misspecified Confounder ($Z_{C2}$) & 0.99 & 0.21 & 0.17 & -0.01 & 0.21 & 0.87 \\ 
\midrule\addlinespace[2.5pt]
\multicolumn{7}{>{\raggedright\arraybackslash}m{1\linewidth}}{Missing Confounder ($Z_{C1}$) - n = 1000} \\[2.5pt] 
\midrule\addlinespace[2.5pt]
Correct Imputation Model & 0.98 & 0.16 & 0.14 & -0.02 & 0.16 & 0.88 \\ 
Oversaturated Imputation Model* & 1.01 & 0.17 & 0.14 & 0.01 & 0.17 & 0.88 \\ 
Misspecified Confounder ($Z_{C2}$) & 0.99 & 0.16 & 0.14 & -0.01 & 0.16 & 0.90 \\ 
\midrule\addlinespace[2.5pt]
\multicolumn{7}{>{\raggedright\arraybackslash}m{1\linewidth}}{Missing Confounder ($Z_{C1}$) - n = 2000} \\[2.5pt] 
\midrule\addlinespace[2.5pt]
Correct Imputation Model & 0.97 & 0.14 & 0.10 & -0.03 & 0.14 & 0.85 \\ 
Oversaturated Imputation Model* & 1.00 & 0.14 & 0.11 & 0.00 & 0.14 & 0.85 \\ 
Misspecified Confounder ($Z_{C2}$) & 0.99 & 0.14 & 0.11 & -0.01 & 0.14 & 0.88 \\ 
\midrule\addlinespace[2.5pt]
\multicolumn{7}{>{\raggedright\arraybackslash}m{1\linewidth}}{Missing Outcome (Y) - n = 100} \\[2.5pt] 
\midrule\addlinespace[2.5pt]
Correct Imputation Model & 1.00 & 0.54 & 0.40 & 0.00 & 0.54 & 0.86 \\ 
Oversaturated Imputation Model* & 0.97 & 0.71 & 0.58 & -0.03 & 0.71 & 0.88 \\ 
Misspecified Confounder ($Z_{C2}$) & 1.02 & 0.45 & 0.34 & 0.02 & 0.45 & 0.85 \\ 
\midrule\addlinespace[2.5pt]
\multicolumn{7}{>{\raggedright\arraybackslash}m{1\linewidth}}{Missing Outcome (Y) - n = 500} \\[2.5pt] 
\midrule\addlinespace[2.5pt]
Correct Imputation Model & 1.00 & 0.23 & 0.20 & 0.00 & 0.23 & 0.89 \\ 
Oversaturated Imputation Model* & 1.01 & 0.23 & 0.19 & 0.01 & 0.23 & 0.87 \\ 
Misspecified Confounder ($Z_{C2}$) & 1.01 & 0.21 & 0.18 & 0.01 & 0.21 & 0.89 \\ 
\midrule\addlinespace[2.5pt]
\multicolumn{7}{>{\raggedright\arraybackslash}m{1\linewidth}}{Missing Outcome (Y) - n = 1000} \\[2.5pt] 
\midrule\addlinespace[2.5pt]
Correct Imputation Model & 0.98 & 0.19 & 0.16 & -0.02 & 0.19 & 0.89 \\ 
Oversaturated Imputation Model* & 1.00 & 0.17 & 0.15 & 0.00 & 0.17 & 0.90 \\ 
Misspecified Confounder ($Z_{C2}$) & 0.99 & 0.16 & 0.14 & -0.01 & 0.16 & 0.89 \\ 
\midrule\addlinespace[2.5pt]
\multicolumn{7}{>{\raggedright\arraybackslash}m{1\linewidth}}{Missing Outcome (Y) - n = 2000} \\[2.5pt] 
\midrule\addlinespace[2.5pt]
Correct Imputation Model & 1.00 & 0.14 & 0.12 & 0.00 & 0.14 & 0.91 \\ 
Oversaturated Imputation Model* & 1.00 & 0.14 & 0.11 & 0.00 & 0.14 & 0.89 \\ 
Misspecified Confounder ($Z_{C2}$) & 1.00 & 0.12 & 0.11 & 0.00 & 0.12 & 0.90 \\ 
\bottomrule

\end{longtable}

\begin{minipage}{\linewidth}
Abbreviations: n = Sample size; Est. = Estimate; MC SE = Monte Carlo Standard Error; Avg. SE = Average Estimated Standard Error; 95\% Cov. = 95\% Confidence Interval Coverage.\\
\end{minipage}
\endgroup

\begingroup
\setlength\LTleft{0\linewidth}
\setlength\LTright{0\linewidth}\fontsize{12.0pt}{14.4pt}\selectfont
\setlength{\LTpost}{0mm}

\begin{longtable}{@{\extracolsep{\fill}}>{\raggedright\arraybackslash}p{\dimexpr 0.30\linewidth -2\tabcolsep-1.5\arrayrulewidth}>{\raggedleft\arraybackslash}p{\dimexpr 0.12\linewidth -2\tabcolsep-1.5\arrayrulewidth}>{\raggedleft\arraybackslash}p{\dimexpr 0.12\linewidth -2\tabcolsep-1.5\arrayrulewidth}>{\raggedleft\arraybackslash}p{\dimexpr 0.12\linewidth -2\tabcolsep-1.5\arrayrulewidth}>{\raggedleft\arraybackslash}p{\dimexpr 0.12\linewidth -2\tabcolsep-1.5\arrayrulewidth}>{\raggedleft\arraybackslash}p{\dimexpr 0.12\linewidth -2\tabcolsep-1.5\arrayrulewidth}>{\raggedleft\arraybackslash}p{\dimexpr 0.12\linewidth -2\tabcolsep-1.5\arrayrulewidth}}

\caption{\label{tbl-5}Simulation results when the data were generated
with a nonlinear relationship between \(Z_{C2}\) and \(Y\) by imputation
model specification. The true average treatment effect is 1.}

\tabularnewline

\toprule
Model Specification & Est. & MC SE & Avg. SE & Bias & RMSE & 95\% Cov. \\ 
\midrule 
\endfirsthead
\toprule
Model Specification & Est. & MC SE & Avg. SE & Bias & RMSE & 95\% Cov. \\ 
\midrule 
\endhead
\midrule\addlinespace[2.5pt]
\multicolumn{7}{>{\raggedright\arraybackslash}m{1\linewidth}}{Missing Confounder ($Z_{C1}$) - n = 100} \\[2.5pt] 
\midrule\addlinespace[2.5pt]
Correct Imputation Model & 0.96 & 1.73 & 1.21 & -0.04 & 1.73 & 0.82 \\ 
Oversaturated Imputation Model* & 0.77 & 1.96 & 1.29 & -0.23 & 1.97 & 0.80 \\ 
Misspecified Confounder ($Z_{C2}$) & 0.47 & 1.91 & 1.36 & -0.53 & 1.98 & 0.82 \\ 
\midrule\addlinespace[2.5pt]
\multicolumn{7}{>{\raggedright\arraybackslash}m{1\linewidth}}{Missing Confounder ($Z_{C1}$) - n = 500} \\[2.5pt] 
\midrule\addlinespace[2.5pt]
Correct Imputation Model & 0.97 & 0.72 & 0.56 & -0.03 & 0.72 & 0.84 \\ 
Oversaturated Imputation Model* & 0.96 & 0.72 & 0.60 & -0.04 & 0.72 & 0.86 \\ 
Misspecified Confounder ($Z_{C2}$) & 0.41 & 0.93 & 0.78 & -0.59 & 1.10 & 0.82 \\ 
\midrule\addlinespace[2.5pt]
\multicolumn{7}{>{\raggedright\arraybackslash}m{1\linewidth}}{Missing Confounder ($Z_{C1}$) - n = 1000} \\[2.5pt] 
\midrule\addlinespace[2.5pt]
Correct Imputation Model & 1.04 & 0.52 & 0.44 & 0.04 & 0.52 & 0.88 \\ 
Oversaturated Imputation Model* & 0.97 & 0.57 & 0.46 & -0.03 & 0.57 & 0.88 \\ 
Misspecified Confounder ($Z_{C2}$) & 0.37 & 0.84 & 0.69 & -0.63 & 1.05 & 0.84 \\ 
\midrule\addlinespace[2.5pt]
\multicolumn{7}{>{\raggedright\arraybackslash}m{1\linewidth}}{Missing Confounder ($Z_{C1}$) - n = 2000} \\[2.5pt] 
\midrule\addlinespace[2.5pt]
Correct Imputation Model & 1.02 & 0.37 & 0.32 & 0.02 & 0.37 & 0.90 \\ 
Oversaturated Imputation Model* & 0.98 & 0.40 & 0.35 & -0.02 & 0.40 & 0.89 \\ 
Misspecified Confounder ($Z_{C2}$) & 0.32 & 0.67 & 0.60 & -0.68 & 0.95 & 0.85 \\ 
\midrule\addlinespace[2.5pt]
\multicolumn{7}{>{\raggedright\arraybackslash}m{1\linewidth}}{Missing Outcome (Y) - n = 100} \\[2.5pt] 
\midrule\addlinespace[2.5pt]
Correct Imputation Model & 1.10 & 2.02 & 1.36 & 0.10 & 2.02 & 0.81 \\ 
Oversaturated Imputation Model* & 1.02 & 2.86 & 1.90 & 0.02 & 2.86 & 0.84 \\ 
Misspecified Confounder ($Z_{C2}$) & -0.15 & 2.59 & 1.71 & -1.15 & 2.83 & 0.79 \\ 
\midrule\addlinespace[2.5pt]
\multicolumn{7}{>{\raggedright\arraybackslash}m{1\linewidth}}{Missing Outcome (Y) - n = 500} \\[2.5pt] 
\midrule\addlinespace[2.5pt]
Correct Imputation Model & 1.04 & 0.77 & 0.60 & 0.04 & 0.77 & 0.85 \\ 
Oversaturated Imputation Model* & 0.86 & 0.85 & 0.64 & -0.14 & 0.86 & 0.84 \\ 
Misspecified Confounder ($Z_{C2}$) & 0.02 & 1.45 & 1.00 & -0.98 & 1.75 & 0.82 \\ 
\midrule\addlinespace[2.5pt]
\multicolumn{7}{>{\raggedright\arraybackslash}m{1\linewidth}}{Missing Outcome (Y) - n = 1000} \\[2.5pt] 
\midrule\addlinespace[2.5pt]
Correct Imputation Model & 1.02 & 0.52 & 0.44 & 0.02 & 0.52 & 0.89 \\ 
Oversaturated Imputation Model* & 0.94 & 0.58 & 0.47 & -0.06 & 0.59 & 0.88 \\ 
Misspecified Confounder ($Z_{C2}$) & 0.05 & 1.12 & 0.82 & -0.95 & 1.47 & 0.77 \\ 
\midrule\addlinespace[2.5pt]
\multicolumn{7}{>{\raggedright\arraybackslash}m{1\linewidth}}{Missing Outcome (Y) - n = 2000} \\[2.5pt] 
\midrule\addlinespace[2.5pt]
Correct Imputation Model & 0.98 & 0.40 & 0.34 & -0.02 & 0.40 & 0.89 \\ 
Oversaturated Imputation Model* & 0.95 & 0.44 & 0.36 & -0.05 & 0.44 & 0.89 \\ 
Misspecified Confounder ($Z_{C2}$) & -0.03 & 0.97 & 0.69 & -1.03 & 1.42 & 0.69 \\ 
\bottomrule

\end{longtable}

\begin{minipage}{\linewidth}
Abbreviations: n = Sample size; Est. = Estimate; MC SE = Monte Carlo Standard Error; Avg. SE = Average Estimated Standard Error; 95\% Cov. = 95\% Confidence Interval Coverage.\\
\end{minipage}
\endgroup

\begingroup
\setlength\LTleft{0\linewidth}
\setlength\LTright{0\linewidth}\fontsize{12.0pt}{14.4pt}\selectfont
\setlength{\LTpost}{0mm}

\begin{longtable}{@{\extracolsep{\fill}}>{\raggedright\arraybackslash}p{\dimexpr 0.30\linewidth -2\tabcolsep-1.5\arrayrulewidth}>{\raggedleft\arraybackslash}p{\dimexpr 0.12\linewidth -2\tabcolsep-1.5\arrayrulewidth}>{\raggedleft\arraybackslash}p{\dimexpr 0.12\linewidth -2\tabcolsep-1.5\arrayrulewidth}>{\raggedleft\arraybackslash}p{\dimexpr 0.12\linewidth -2\tabcolsep-1.5\arrayrulewidth}>{\raggedleft\arraybackslash}p{\dimexpr 0.12\linewidth -2\tabcolsep-1.5\arrayrulewidth}>{\raggedleft\arraybackslash}p{\dimexpr 0.12\linewidth -2\tabcolsep-1.5\arrayrulewidth}>{\raggedleft\arraybackslash}p{\dimexpr 0.12\linewidth -2\tabcolsep-1.5\arrayrulewidth}}

\caption{\label{tbl-6}Simulation results when the data were generated
with a nonlinear relationship between both \(Z_{C2}\) and \(X\) and
\(Z_{C2}\) and \(Y\) by imputation model specification. The true average
treatment effect is 1.}

\tabularnewline

\toprule
Model Specification & Est. & MC SE & Avg. SE & Bias & RMSE & 95\% Cov. \\ 
\midrule 
\endfirsthead
\toprule
Model Specification & Est. & MC SE & Avg. SE & Bias & RMSE & 95\% Cov. \\ 
\midrule 
\endhead
\midrule\addlinespace[2.5pt]
\multicolumn{7}{>{\raggedright\arraybackslash}m{1\linewidth}}{Missing Confounder ($Z_{C1}$) - n = 100} \\[2.5pt] 
\midrule\addlinespace[2.5pt]
Correct Imputation Model & 1.13 & 3.78 & 2.43 & 0.13 & 3.78 & 0.75 \\ 
Oversaturated Imputation Model* & 0.66 & 4.07 & 2.61 & -0.34 & 4.08 & 0.74 \\ 
Misspecified Confounder ($Z_{C2}$) & 0.31 & 3.65 & 2.54 & -0.69 & 3.71 & 0.76 \\ 
\midrule\addlinespace[2.5pt]
\multicolumn{7}{>{\raggedright\arraybackslash}m{1\linewidth}}{Missing Confounder ($Z_{C1}$) - n = 500} \\[2.5pt] 
\midrule\addlinespace[2.5pt]
Correct Imputation Model & 1.12 & 1.65 & 1.16 & 0.12 & 1.66 & 0.78 \\ 
Oversaturated Imputation Model* & 0.83 & 1.77 & 1.21 & -0.17 & 1.78 & 0.77 \\ 
Misspecified Confounder ($Z_{C2}$) & 0.31 & 1.82 & 1.43 & -0.69 & 1.95 & 0.80 \\ 
\midrule\addlinespace[2.5pt]
\multicolumn{7}{>{\raggedright\arraybackslash}m{1\linewidth}}{Missing Confounder ($Z_{C1}$) - n = 1000} \\[2.5pt] 
\midrule\addlinespace[2.5pt]
Correct Imputation Model & 1.02 & 1.36 & 0.95 & 0.02 & 1.36 & 0.78 \\ 
Oversaturated Imputation Model* & 1.02 & 1.37 & 0.99 & 0.02 & 1.36 & 0.80 \\ 
Misspecified Confounder ($Z_{C2}$) & 0.11 & 1.58 & 1.25 & -0.89 & 1.81 & 0.79 \\ 
\midrule\addlinespace[2.5pt]
\multicolumn{7}{>{\raggedright\arraybackslash}m{1\linewidth}}{Missing Confounder ($Z_{C1}$) - n = 2000} \\[2.5pt] 
\midrule\addlinespace[2.5pt]
Correct Imputation Model & 1.06 & 1.10 & 0.73 & 0.06 & 1.10 & 0.80 \\ 
Oversaturated Imputation Model* & 1.00 & 1.06 & 0.78 & 0.00 & 1.06 & 0.80 \\ 
Misspecified Confounder ($Z_{C2}$) & 0.15 & 1.40 & 1.03 & -0.85 & 1.64 & 0.80 \\ 
\midrule\addlinespace[2.5pt]
\multicolumn{7}{>{\raggedright\arraybackslash}m{1\linewidth}}{Missing Outcome (Y) - n = 100} \\[2.5pt] 
\midrule\addlinespace[2.5pt]
Correct Imputation Model & 1.34 & 4.79 & 2.98 & 0.34 & 4.80 & 0.76 \\ 
Oversaturated Imputation Model* & 0.43 & 8.14 & 5.62 & -0.57 & 8.15 & 0.81 \\ 
Misspecified Confounder ($Z_{C2}$) & -0.01 & 4.44 & 2.96 & -1.01 & 4.55 & 0.77 \\ 
\midrule\addlinespace[2.5pt]
\multicolumn{7}{>{\raggedright\arraybackslash}m{1\linewidth}}{Missing Outcome (Y) - n = 500} \\[2.5pt] 
\midrule\addlinespace[2.5pt]
Correct Imputation Model & 1.10 & 1.79 & 1.36 & 0.10 & 1.79 & 0.79 \\ 
Oversaturated Imputation Model* & 1.03 & 1.91 & 1.43 & 0.03 & 1.91 & 0.79 \\ 
Misspecified Confounder ($Z_{C2}$) & -0.16 & 2.34 & 1.58 & -1.16 & 2.61 & 0.78 \\ 
\midrule\addlinespace[2.5pt]
\multicolumn{7}{>{\raggedright\arraybackslash}m{1\linewidth}}{Missing Outcome (Y) - n = 1000} \\[2.5pt] 
\midrule\addlinespace[2.5pt]
Correct Imputation Model & 0.89 & 1.38 & 1.03 & -0.11 & 1.38 & 0.80 \\ 
Oversaturated Imputation Model* & 0.91 & 1.37 & 1.07 & -0.09 & 1.37 & 0.80 \\ 
Misspecified Confounder ($Z_{C2}$) & -0.23 & 2.06 & 1.26 & -1.23 & 2.40 & 0.72 \\ 
\midrule\addlinespace[2.5pt]
\multicolumn{7}{>{\raggedright\arraybackslash}m{1\linewidth}}{Missing Outcome (Y) - n = 2000} \\[2.5pt] 
\midrule\addlinespace[2.5pt]
Correct Imputation Model & 1.02 & 1.01 & 0.79 & 0.02 & 1.01 & 0.83 \\ 
Oversaturated Imputation Model* & 0.85 & 1.12 & 0.82 & -0.15 & 1.13 & 0.81 \\ 
Misspecified Confounder ($Z_{C2}$) & 0.06 & 1.67 & 1.04 & -0.94 & 1.92 & 0.76 \\ 
\bottomrule

\end{longtable}

\begin{minipage}{\linewidth}
Abbreviations: n = Sample size; Est. = Estimate; MC SE = Monte Carlo Standard Error; Avg. SE = Average Estimated Standard Error; 95\% Cov. = 95\% Confidence Interval Coverage.\\
\end{minipage}
\endgroup

\section{Implementation}\label{implementation}

To illustrate the recommended workflow, here we walk through the
nonlinear case with treatment heterogeneity used in the simulation study
(Figure~\ref{fig-schema}). Below we (i) generate one dataset under the
scenario described in Section~\ref{sec-sim-1}, (ii) induce missingness
in the confounder, (iii) impute the confounder separately by exposure to
preserve congeniality with an outcome model that includes an
interaction, (iv) estimate the causal effect in each imputed dataset
using IPW and a nonlinear outcome model, and (v) pool estimates using
Rubin's rules. We set \(m=20\) imputations throughout.

For the analysis we created a helper \texttt{fit\_ipw\_effect()} which
fits propensity score weights (using \texttt{weightit} from the WeightIt
package (Greifer 2025)), fits the weighted outcome model (using
\texttt{lm\_weightit} from the WeightIt package), and extracts the
average treatment effect via \texttt{avg\_comparisons()} from the
marginaleffects package (Arel-Bundock, Greifer, and Heiss 2024).

\begin{Shaded}
\begin{Highlighting}[]
\FunctionTok{library}\NormalTok{(WeightIt)}
\FunctionTok{library}\NormalTok{(marginaleffects)}
\NormalTok{fit\_ipw\_effect }\OtherTok{\textless{}{-}} \ControlFlowTok{function}\NormalTok{(.data, .x, .ps\_fmla, .outcome\_fmla) \{}
\NormalTok{  w }\OtherTok{\textless{}{-}} \FunctionTok{weightit}\NormalTok{(.ps\_fmla, }\AttributeTok{data =}\NormalTok{ .data)}
\NormalTok{  o }\OtherTok{\textless{}{-}} \FunctionTok{lm\_weightit}\NormalTok{(.outcome\_fmla, }\AttributeTok{data =}\NormalTok{ .data, }\AttributeTok{weightit =}\NormalTok{ w)}
\NormalTok{  e }\OtherTok{\textless{}{-}} \FunctionTok{avg\_comparisons}\NormalTok{(o, }\AttributeTok{variables =}\NormalTok{ .x) }
  \FunctionTok{return}\NormalTok{(}\FunctionTok{c}\NormalTok{(}\AttributeTok{estimate =}\NormalTok{ e}\SpecialCharTok{$}\NormalTok{estimate, }\AttributeTok{std.err =}\NormalTok{ e}\SpecialCharTok{$}\NormalTok{std.error))}
\NormalTok{\}}
\end{Highlighting}
\end{Shaded}

\subsection{Data generation}\label{data-generation}

We simulate the same data-generating mechanism used in the
Section~\ref{sec-sim-1}: three continuous covariates \(Z_I\)
(\texttt{zi}), \(Z_P\) (\texttt{zp}) and \(Z_C\) (\texttt{zc}), a binary
exposure \(X\) and an outcome that is nonlinear in \(Z_P\) and includes
an \(X\times Z_C\) interaction. Missingness is induced in \(Z_C\) with
probability depending on \(X\) to achieve approximately 20\%
missingness.

\begin{Shaded}
\begin{Highlighting}[]
\FunctionTok{set.seed}\NormalTok{(}\DecValTok{928}\NormalTok{)}
\NormalTok{n }\OtherTok{\textless{}{-}} \DecValTok{2000}

\NormalTok{zp }\OtherTok{\textless{}{-}} \FunctionTok{rnorm}\NormalTok{(n)          }
\NormalTok{zi }\OtherTok{\textless{}{-}} \FunctionTok{rnorm}\NormalTok{(n)         }
\NormalTok{zc  }\OtherTok{\textless{}{-}} \FunctionTok{rnorm}\NormalTok{(n, }\DecValTok{1}\NormalTok{) }

\NormalTok{x }\OtherTok{\textless{}{-}} \FunctionTok{rbinom}\NormalTok{(n, }\DecValTok{1}\NormalTok{, }\FunctionTok{plogis}\NormalTok{(}\DecValTok{1} \SpecialCharTok{{-}}\NormalTok{ zc }\SpecialCharTok{+} \DecValTok{2} \SpecialCharTok{*}\NormalTok{ zi))}
\NormalTok{y }\OtherTok{\textless{}{-}} \FloatTok{0.5} \SpecialCharTok{*}\NormalTok{ x }\SpecialCharTok{+}\NormalTok{ zc }\SpecialCharTok{+} \DecValTok{2} \SpecialCharTok{*}\NormalTok{ (zp}\SpecialCharTok{\^{}}\DecValTok{2}\NormalTok{) }\SpecialCharTok{+} \FloatTok{0.5} \SpecialCharTok{*}\NormalTok{ x }\SpecialCharTok{*}\NormalTok{ zc }\SpecialCharTok{+} \FunctionTok{rnorm}\NormalTok{(n)}

\NormalTok{p\_missing }\OtherTok{\textless{}{-}} \FunctionTok{plogis}\NormalTok{(}\SpecialCharTok{{-}}\FloatTok{1.15} \SpecialCharTok{{-}} \FloatTok{0.5} \SpecialCharTok{*}\NormalTok{ x)}
\NormalTok{zc\_obs }\OtherTok{\textless{}{-}} \FunctionTok{ifelse}\NormalTok{(}\FunctionTok{rbinom}\NormalTok{(n, }\DecValTok{1}\NormalTok{, p\_missing) }\SpecialCharTok{==} \DecValTok{1}\NormalTok{, }\ConstantTok{NA}\NormalTok{, zc)}

\NormalTok{data }\OtherTok{\textless{}{-}} \FunctionTok{data.frame}\NormalTok{(}\AttributeTok{x\_obs =}\NormalTok{ x, }
                   \AttributeTok{y\_obs =}\NormalTok{ y, }
                   \AttributeTok{zc\_obs =}\NormalTok{ zc\_obs, }
                   \AttributeTok{zi =}\NormalTok{ zi, }
                   \AttributeTok{zp =}\NormalTok{ zp)}
\end{Highlighting}
\end{Shaded}

\subsection{Multiple imputation (separate by
exposure)}\label{multiple-imputation-separate-by-exposure}

To ensure congeniality with the outcome model that includes an
\(X\times Z_C\) interaction, we impute \texttt{zc\_obs} separately
within the exposed and unexposed strata. For simplicity, we load the
tidyverse package (Wickham et al. 2019) for data manipulation (filtering
to split the data by the exposure) and recombining the completed
datasets (with \texttt{map}). The imputation model includes the outcome
\texttt{y\_obs} and a spline term for the precision variable \texttt{zp}
to capture nonlinearity. The \texttt{make.formulas} function from the
\texttt{mice} package is a helper function that creates formulas for
each of the variables in a data frame. By default, it includes all other
variables present in the data frame in an additive model (with no
nonlinear terms or interactions). We can update components of this
object as needed, for example here we will update to include the
nonlinear term for \texttt{zp}. We then use the \texttt{mice} function
to perform the imputations and \texttt{complete} to create the \(m=20\)
complete datasets.

\begin{Shaded}
\begin{Highlighting}[]
\FunctionTok{library}\NormalTok{(tidyverse)}
\FunctionTok{library}\NormalTok{(mice)}
\end{Highlighting}
\end{Shaded}

\begin{Shaded}
\begin{Highlighting}[]
\NormalTok{m }\OtherTok{\textless{}{-}} \DecValTok{20}

\CommentTok{\# split by exposure}
\NormalTok{data1 }\OtherTok{\textless{}{-}} \FunctionTok{filter}\NormalTok{(data, x\_obs }\SpecialCharTok{==} \DecValTok{1}\NormalTok{)}
\NormalTok{data0 }\OtherTok{\textless{}{-}} \FunctionTok{filter}\NormalTok{(data, x\_obs }\SpecialCharTok{==} \DecValTok{0}\NormalTok{)}

\CommentTok{\# update imputation formulas to include spline}
\NormalTok{f }\OtherTok{\textless{}{-}} \FunctionTok{make.formulas}\NormalTok{(data)}
\NormalTok{f}\SpecialCharTok{$}\NormalTok{zc\_obs }\OtherTok{\textless{}{-}}\NormalTok{ zc\_obs }\SpecialCharTok{\textasciitilde{}}\NormalTok{ y\_obs }\SpecialCharTok{+}\NormalTok{ splines}\SpecialCharTok{::}\FunctionTok{ns}\NormalTok{(zp, }\DecValTok{3}\NormalTok{)}

\NormalTok{imp1 }\OtherTok{\textless{}{-}} \FunctionTok{mice}\NormalTok{(data1,}
             \AttributeTok{method =} \StringTok{"norm"}\NormalTok{,}
             \AttributeTok{formulas =}\NormalTok{ f, }
             \AttributeTok{m =}\NormalTok{ m, }
             \AttributeTok{print =} \ConstantTok{FALSE}\NormalTok{)}
\end{Highlighting}
\end{Shaded}

\begin{Shaded}
\begin{Highlighting}[]
\NormalTok{imp0 }\OtherTok{\textless{}{-}} \FunctionTok{mice}\NormalTok{(data0,}
             \AttributeTok{method =} \StringTok{"norm"}\NormalTok{, }
             \AttributeTok{formulas =}\NormalTok{ f, }
             \AttributeTok{m =}\NormalTok{ m, }
             \AttributeTok{print =} \ConstantTok{FALSE}\NormalTok{)}
\end{Highlighting}
\end{Shaded}

\begin{Shaded}
\begin{Highlighting}[]
\CommentTok{\# reassemble 20 full datasets }
\NormalTok{completed\_list }\OtherTok{\textless{}{-}} \FunctionTok{map}\NormalTok{(}\DecValTok{1}\SpecialCharTok{:}\NormalTok{m,}
                      \SpecialCharTok{\textasciitilde{}} \FunctionTok{bind\_rows}\NormalTok{(}\FunctionTok{complete}\NormalTok{(imp1, .x),}
                                  \FunctionTok{complete}\NormalTok{(imp0, .x)))}
\end{Highlighting}
\end{Shaded}

\subsection{Analysis within each imputed
dataset}\label{analysis-within-each-imputed-dataset}

For each completed dataset we estimate propensity score weights with
\texttt{x\_obs\ \textasciitilde{}\ zc\_obs} and then fit the weighted
nonlinear outcome regression that includes a cubic spline for
\texttt{zp}, and the interaction \texttt{x\_obs\ :\ zc\_obs}. The ATE is
extracted via the \texttt{fit\_ipw\_effect()} helper function that we
created above. This will output an estimate and standard error for each
of the 20 imputed datasets.

\begin{Shaded}
\begin{Highlighting}[]
\NormalTok{results }\OtherTok{\textless{}{-}} \FunctionTok{map\_dfr}\NormalTok{(completed\_list, }\SpecialCharTok{\textasciitilde{}}\NormalTok{ \{}
\NormalTok{  v }\OtherTok{\textless{}{-}} \FunctionTok{fit\_ipw\_effect}\NormalTok{(}
\NormalTok{    .x,}
    \AttributeTok{.x =} \StringTok{"x\_obs"}\NormalTok{,}
    \AttributeTok{.ps\_fmla =}\NormalTok{ x\_obs }\SpecialCharTok{\textasciitilde{}}\NormalTok{ zc\_obs,}
    \AttributeTok{.outcome\_fmla =}\NormalTok{ y\_obs }\SpecialCharTok{\textasciitilde{}}\NormalTok{ x\_obs }\SpecialCharTok{+}\NormalTok{ zc\_obs }\SpecialCharTok{+} 
\NormalTok{      splines}\SpecialCharTok{::}\FunctionTok{ns}\NormalTok{(zp, }\DecValTok{3}\NormalTok{) }\SpecialCharTok{+}\NormalTok{ x\_obs}\SpecialCharTok{:}\NormalTok{zc\_obs)}
  \FunctionTok{tibble}\NormalTok{(}\AttributeTok{estimate =} \FunctionTok{as.numeric}\NormalTok{(v[}\StringTok{"estimate"}\NormalTok{]),}
         \AttributeTok{std.err =} \FunctionTok{as.numeric}\NormalTok{(v[}\StringTok{"std.err"}\NormalTok{]))}
\NormalTok{\})}
\end{Highlighting}
\end{Shaded}

\subsection{Pooling (Rubin's rules)}\label{pooling-rubins-rules}

We combine the point estimates \(\hat{\Delta}^{(j)}\) and their
within-imputation variances \(U^{(j)}\) across the \(m\) completed
datasets using Rubin's rules. The pooled estimate is
\(\bar{\Delta} = m^{-1}\sum_j \hat{\Delta}^{(j)}\) and the total
variance is \(T = \bar{U} + (1 + m^{-1})B\) (as described in
Section~\ref{sec-mi}).

\begin{Shaded}
\begin{Highlighting}[]
\NormalTok{q\_bar }\OtherTok{\textless{}{-}} \FunctionTok{mean}\NormalTok{(results}\SpecialCharTok{$}\NormalTok{estimate)  }\CommentTok{\# pooled point estimate}
\NormalTok{u\_bar }\OtherTok{\textless{}{-}} \FunctionTok{mean}\NormalTok{(results}\SpecialCharTok{$}\NormalTok{std.err}\SpecialCharTok{\^{}}\DecValTok{2}\NormalTok{) }\CommentTok{\# avg within{-}imputation variance}
\NormalTok{b     }\OtherTok{\textless{}{-}} \FunctionTok{var}\NormalTok{(results}\SpecialCharTok{$}\NormalTok{estimate)   }\CommentTok{\# between{-}imputation variance}
\NormalTok{t\_var }\OtherTok{\textless{}{-}}\NormalTok{ u\_bar }\SpecialCharTok{+}\NormalTok{ (}\DecValTok{1} \SpecialCharTok{+} \DecValTok{1}\SpecialCharTok{/}\NormalTok{m) }\SpecialCharTok{*}\NormalTok{ b}
\NormalTok{pooled\_se }\OtherTok{\textless{}{-}} \FunctionTok{sqrt}\NormalTok{(t\_var)}

\FunctionTok{tibble}\NormalTok{(}\AttributeTok{estimate =}\NormalTok{ q\_bar, }\AttributeTok{std.error =}\NormalTok{ pooled\_se)}
\end{Highlighting}
\end{Shaded}

\begin{verbatim}
# A tibble: 1 x 2
  estimate std.error
     <dbl>     <dbl>
1    0.984    0.0549
\end{verbatim}

We can now report the \texttt{estimate} and \texttt{std.error} above as
the pooled ATE and its standard error.

\section{Conclusion}\label{conclusion}

This study provides clear guidance on the proper specification of
imputation models when combining multiple imputation with propensity
score methods for causal inference. A central takeaway is the critical
importance of representing variables in their correct functional form
within the imputation model, a detail often overlooked in practice.
While commonly used software such as mice (van Buuren and
Groothuis-Oudshoorn 2011) include all variables by default (a good
practice), the default is to include them linearly, which can introduce
bias when the relationships in analysis models are treated as nonlinear.

Our findings demonstrate that the imputation model must include all
variables used in either the propensity score model or the outcome
model, not just in name, but in the same functional form as in the
analysis models. This ensures congeniality and supports valid estimation
of treatment effects. The separate patterns observed when either
confounders (\(Z_C\)) or outcomes (\(Y\)) are missing highlight the need
for careful consideration based on the specific missing data scenario.

An important challenge for practitioners is that there may not be a
shared understanding of what is meant by including ``all'' variables in
the imputation model. For some researchers, ``all'' may refer only to
confounders, excluding the exposure or outcome variables. Others may
interpret it as including every variable in the dataset but only in
linear form, regardless of how they are specified in the analysis
models. Still others may consider only main effects as sufficient,
overlooking interaction terms or nonlinear transformations that appear
in their analysis models. This lack of consensus around what constitutes
a complete imputation model specification can lead to seemingly
well-intentioned but ultimately inadequate implementations. The issue is
compounded by the fact that many practitioners may believe they are
following best practices by including ``all variables'' without
recognizing that their interpretation of completeness differs from what
congeniality actually requires. Establishing clearer operational
definitions of comprehensive imputation model specification, one that
explicitly encompasses exposure variables, outcome variables,
appropriate functional forms, and relevant interactions, is essential
for moving the field toward more consistent and valid practice.

While the principle of congeniality is well established within the
missing data methodological community, its importance may not always be
fully recognized or consistently applied across different areas of
research, particularly in scenarios where multiple analysis models are
used. These principles of careful imputation model specification apply
broadly across different types of analyses involving multiple models,
not only in causal inference settings. Whether the research goal is
prediction, description, or causal explanation, ensuring that imputation
models adequately reflect the complexity and functional forms used in
subsequent analyses remains critical for valid statistical inference.
Our theoretical and simulated results show that excluding variables from
the imputation model, even those that are not confounders (e.g.,
precision variables), can introduce substantial bias in treatment effect
estimates when confounders are imputed. This bias remains even when
using doubly robust estimators (and potentially worsens, as this
introduces an additional model that the imputation model must be
congenial with).

In scenarios with nonlinear relationships, we found that misspecifying
the functional form of a variable in the imputation model led to bias
similar in magnitude to excluding the variable entirely. This result
underscores the need for careful alignment between the imputation model
and the analysis models. Our findings regarding treatment effect
heterogeneity further emphasize this point, as we demonstrated that
separate imputation models for exposed and unexposed groups may be
necessary when the outcome is missing and the true data-generating
process includes treatment effect heterogeneity.

The complete case analysis results offer additional nuance to our
understanding, showing that complete case analysis can be appropriate in
some circumstances (when confounders are missing under specific
conditions) but biased in others (when the outcome is missing), even
with missing at random mechanisms. This highlights the complex interplay
between causal structure and missing data mechanisms.

While this work focused on continuous outcomes with a binary treatment,
similar principles apply to other outcome types and data structures.
Time-to-event outcomes present additional complexities for multiple
imputation, as the imputation model must appropriately handle censoring
mechanisms and potentially time-varying effects. The bias from
imputation model misspecification may be even more pronounced in
survival settings due to the challenges of adequately representing the
hazard function and its relationship with covariates in the imputation
model. In addition, the impact of incorporating modern machine learning
techniques in both the model fitting and imputation steps represents an
important direction for future work.

The key practical recommendations from our study are:

\begin{enumerate}
\def\labelenumi{\arabic{enumi}.}
\item
  Include all variables from the propensity score or outcome models in
  the imputation model, regardless of their causal role (confounders or
  precision variables), along with the exposure (when imputing a
  confounder or the outcome) and the outcome (when imputing a
  confounder).
\item
  Ensure that these variables are included in the same functional form
  as they appear in the analysis models (e.g., if quadratic terms are
  used in the analysis, they should also be included in the imputation).
\item
  When treatment effect heterogeneity is accounted for in the analysis
  modeling procedure, fit separate imputation models by treatment
  status.
\item
  Understand the relationship between missing data mechanisms and causal
  structure when deciding whether complete case analysis is appropriate.
\item
  When in doubt, err on the side of inclusion rather than exclusion when
  specifying the imputation model.
\end{enumerate}

These recommendations align with fundamental principles of proper
multiple imputation. The imputation model must be at least as rich as
the analysis model to preserve the joint distribution of all variables
relevant to the analysis. Our study extends these principles to the
specific context of causal inference where multiple models (propensity
score and outcome) may be involved. By following these guidelines,
researchers can more confidently apply multiple imputation in
observational studies where both missing data and confounding are
present, leading to more valid and reliable estimates of causal effects.

\section*{References}\label{references}
\addcontentsline{toc}{section}{References}

\phantomsection\label{refs}


\begin{CSLReferences}{1}{0}
\bibitem[\citeproctext]{ref-marginaleffects}
Arel-Bundock, Vincent, Noah Greifer, and Andrew Heiss. 2024. {``How to
Interpret Statistical Models Using {marginaleffects} for {R} and
{Python}.''} \emph{Journal of Statistical Software} 111 (9): 1--32.
\url{https://doi.org/10.18637/jss.v111.i09}.

\bibitem[\citeproctext]{ref-bang2005doubly}
Bang, Heejung, and James M Robins. 2005. {``Doubly Robust Estimation in
Missing Data and Causal Inference Models.''} \emph{Biometrics} 61 (4):
962--73.

\bibitem[\citeproctext]{ref-bartlett2020bootstrap}
Bartlett, Jonathan W, and Rachael A Hughes. 2020. {``Bootstrap Inference
for Multiple Imputation Under Uncongeniality and Misspecification.''}
\emph{Statistical Methods in Medical Research} 29 (12): 3533--46.

\bibitem[\citeproctext]{ref-brookhart2006variable}
Brookhart, M Alan, Sebastian Schneeweiss, Kenneth J Rothman, Robert J
Glynn, Jerry Avorn, and Til Stürmer. 2006. {``Variable Selection for
Propensity Score Models.''} \emph{American Journal of Epidemiology} 163
(12): 1149--56.

\bibitem[\citeproctext]{ref-craycroft2020propensity}
Craycroft, John A, Jiapeng Huang, and Maiying Kong. 2020. {``Propensity
Score Specification for Optimal Estimation of Average Treatment Effect
with Binary Response.''} \emph{Statistical Methods in Medical Research}
29 (12): 3623--40.

\bibitem[\citeproctext]{ref-d2024behind}
D'Agostino McGowan, Lucy, Sarah C Lotspeich, and Staci A Hepler. 2024.
{``The "Why" Behind Including "y" in Your Imputation Model.''}
\emph{Statistical Methods in Medical Research} 33 (6): 996--1020.

\bibitem[\citeproctext]{ref-daniels2014fully}
Daniels, MJ, Chenguang Wang, and BH Marcus. 2014. {``Fully Bayesian
Inference Under Ignorable Missingness in the Presence of Auxiliary
Covariates.''} \emph{Biometrics} 70 (1): 62--72.

\bibitem[\citeproctext]{ref-dashti2024handling}
Dashti, S Ghazaleh, Katherine J Lee, Julie A Simpson, Ian R White, John
B Carlin, and Margarita Moreno-Betancur. 2024. {``Handling Missing Data
When Estimating Causal Effects with Targeted Maximum Likelihood
Estimation.''} \emph{American Journal of Epidemiology} 193 (7):
1019--30.

\bibitem[\citeproctext]{ref-de2011covariate}
De Luna, Xavier, Ingeborg Waernbaum, and Thomas S Richardson. 2011.
{``Covariate Selection for the Nonparametric Estimation of an Average
Treatment Effect.''} \emph{Biometrika} 98 (4): 861--75.

\bibitem[\citeproctext]{ref-franklin2015regularized}
Franklin, Jessica M, Wesley Eddings, Robert J Glynn, and Sebastian
Schneeweiss. 2015. {``Regularized Regression Versus the High-Dimensional
Propensity Score for Confounding Adjustment in Secondary Database
Analyses.''} \emph{American Journal of Epidemiology} 182 (7): 651--59.

\bibitem[\citeproctext]{ref-granger2019avoiding}
Granger, Emily, Jamie C Sergeant, and Mark Lunt. 2019. {``Avoiding
Pitfalls When Combining Multiple Imputation and Propensity Scores.''}
\emph{Statistics in Medicine} 38 (26): 5120--32.

\bibitem[\citeproctext]{ref-weightit}
Greifer, Noah. 2025. \emph{WeightIt: Weighting for Covariate Balance in
Observational Studies}.
\url{https://CRAN.R-project.org/package=WeightIt}.

\bibitem[\citeproctext]{ref-hahn2004functional}
Hahn, Jinyong. 2004. {``Functional Restriction and Efficiency in Causal
Inference.''} \emph{The Review of Economics and Statistics} 86 (1):
73--76.

\bibitem[\citeproctext]{ref-von2013should}
Hippel, Paul T von. 2013. {``Should a Normal Imputation Model Be
Modified to Impute Skewed Variables?''} \emph{Sociological Methods \&
Research} 42 (1): 105--38.

\bibitem[\citeproctext]{ref-kang2007demystifying}
Kang, Joseph DY, and Joseph L Schafer. 2007. {``Demystifying Double
Robustness: A Comparison of Alternative Strategies for Estimating a
Population Mean from Incomplete Data.''} \emph{Statistical Science} 22
(4): 523--39.

\bibitem[\citeproctext]{ref-leyrat2019propensity}
Leyrat, Clémence, Shaun R Seaman, Ian R White, Ian Douglas, Liam Smeeth,
Joseph Kim, Matthieu Resche-Rigon, James R Carpenter, and Elizabeth J
Williamson. 2019. {``Propensity Score Analysis with Partially Observed
Covariates: How Should Multiple Imputation Be Used?''} \emph{Statistical
Methods in Medical Research} 28 (1): 3--19.

\bibitem[\citeproctext]{ref-lunceford2004stratification}
Lunceford, Jared K, and Marie Davidian. 2004. {``Stratification and
Weighting via the Propensity Score in Estimation of Causal Treatment
Effects: A Comparative Study.''} \emph{Statistics in Medicine} 23 (19):
2937--60.

\bibitem[\citeproctext]{ref-mitra2016comparison}
Mitra, Robin, and Jerome P Reiter. 2016. {``A Comparison of Two Methods
of Estimating Propensity Scores After Multiple Imputation.''}
\emph{Statistical Methods in Medical Research} 25 (1): 188--204.

\bibitem[\citeproctext]{ref-penning2016comments}
Penning de Vries, BBL, and Rolf HH Groenwold. 2016. {``Comments on
Propensity Score Matching Following Multiple Imputation.''}
\emph{Statistical Methods in Medical Research}. SAGE Publications Sage
UK: London, England.

\bibitem[\citeproctext]{ref-robins2000robust}
Robins, James M. 2000. {``Robust Estimation in Sequentially Ignorable
Missing Data and Causal Inference Models.''} In \emph{Proceedings of the
American Statistical Association}, 1999:6--10. Indianapolis, IN.

\bibitem[\citeproctext]{ref-robins1994estimation}
Robins, James M, Andrea Rotnitzky, and Lue Ping Zhao. 1994.
{``Estimation of Regression Coefficients When Some Regressors Are Not
Always Observed.''} \emph{Journal of the American Statistical
Association} 89 (427): 846--66.

\bibitem[\citeproctext]{ref-rubin1987}
Rubin, D.B. 1987. \emph{Multiple Imputation for Nonresponse in Surveys}.
New York (NY): J. Wiley \& Sons.

\bibitem[\citeproctext]{ref-scharfstein1999adjusting}
Scharfstein, Daniel O, Andrea Rotnitzky, and James M Robins. 1999.
{``Adjusting for Nonignorable Drop-Out Using Semiparametric Nonresponse
Models.''} \emph{Journal of the American Statistical Association} 94
(448): 1096--1120.

\bibitem[\citeproctext]{ref-tang2023ultra}
Tang, Dingke, Dehan Kong, Wenliang Pan, and Linbo Wang. 2023.
{``Ultra-High Dimensional Variable Selection for Doubly Robust Causal
Inference.''} \emph{Biometrics} 79 (2): 903--14.

\bibitem[\citeproctext]{ref-mice}
van Buuren, Stef, and Karin Groothuis-Oudshoorn. 2011. {``{mice}:
Multivariate Imputation by Chained Equations in r.''} \emph{Journal of
Statistical Software} 45 (3): 1--67.
\url{https://doi.org/10.18637/jss.v045.i03}.

\bibitem[\citeproctext]{ref-van2012flexible}
Van Buuren, Stef, and Stef Van Buuren. 2012. \emph{Flexible Imputation
of Missing Data}. Vol. 10. CRC press Boca Raton, FL.

\bibitem[\citeproctext]{ref-van2011targeted}
Van der Laan, Mark J, Sherri Rose, et al. 2011. \emph{Targeted Learning:
Causal Inference for Observational and Experimental Data}. Vol. 4.
Springer.

\bibitem[\citeproctext]{ref-von20098}
Von Hippel, Paul T. 2009. {``8. How to Impute Interactions, Squares, and
Other Transformed Variables.''} \emph{Sociological Methodology} 39 (1):
265--91.

\bibitem[\citeproctext]{ref-white2011multiple}
White, Ian R, Patrick Royston, and Angela M Wood. 2011. {``Multiple
Imputation Using Chained Equations: Issues and Guidance for Practice.''}
\emph{Statistics in Medicine} 30 (4): 377--99.

\bibitem[\citeproctext]{ref-tidyverse}
Wickham, Hadley, Mara Averick, Jennifer Bryan, Winston Chang, Lucy
D'Agostino McGowan, Romain François, Garrett Grolemund, et al. 2019.
{``Welcome to the {tidyverse}.''} \emph{Journal of Open Source Software}
4 (43): 1686. \url{https://doi.org/10.21105/joss.01686}.

\bibitem[\citeproctext]{ref-williamson2012doubly}
Williamson, Elizabeth Jane, A Forbes, and R Wolfe. 2012. {``Doubly
Robust Estimators of Causal Exposure Effects with Missing Data in the
Outcome, Exposure or a Confounder.''} \emph{Statistics in Medicine} 31
(30): 4382--4400.

\end{CSLReferences}
\end{document}